\documentclass[aps,prd,notitlepage,superscriptaddress,showpacs,preprintnumbers,nofootinbib,10pt,floatfix,twocolumn]{revtex4-1}
\usepackage{amscd}
\usepackage{microtype}
\usepackage{graphicx}
\usepackage{float}
\usepackage{wrapfig}
\usepackage{bm}
\usepackage{color}
\usepackage{comment}
\usepackage{xcolor}
\usepackage[normalem]{ulem}
\usepackage[colorlinks=true,urlcolor=blue,linkcolor=blue,citecolor=blue,hypertexnames]{hyperref}
\usepackage{mathrsfs}
\usepackage{booktabs}
\usepackage{natbib}
\usepackage{dcolumn}
\usepackage{url}
\usepackage{amssymb} 
\usepackage{amsmath} 
\usepackage[noabbrev]{cleveref}
\usepackage{booktabs}
\hypersetup{
colorlinks=true,
citecolor=blue,
citebordercolor=red,
linktoc=all,
linkcolor=blue,
urlcolor=blue
}
\allowdisplaybreaks

\catcode`\@=11
\def\lsim{\mathrel{\mathpalette\@versim<}}
\def\gsim{\mathrel{\mathpalette\@versim>}}
\def\@versim#1#2{\vcenter{\offinterlineskip
\ialign{$\m@th#1\hfil##\hfil$\crcr#2\crcr\sim\crcr } }}
\catcode`\@=12
\newcommand{\Slash}[1]{{\ooalign{\hfil/\hfil\crcr$#1$}}} 
\newcommand{\bvec}[1]{\mbox{\boldmath $#1$}}

\newcommand{\p}{\partial}

\newcommand{\al}[1]{\begin{align}#1\end{align}}
\newcommand{\bp}{\begin{pmatrix}}
\newcommand{\ep}{\end{pmatrix}}
\newcommand{\nn}{\nonumber\\}

\newcommand{\df}{\text{d}}

\newcommand{\bs}[1]{\boldsymbol}

\newcommand{\Tr}{{\rm Tr}\,}
\newcommand{\tr}{{\rm tr}\,}

\newcommand{\pmat}[1]{\begin{pmatrix}#1\end{pmatrix}}

\newcommand{\fn}[1]{\!\left(#1\right)}

\newcommand{\sr}[2]{\stackrel{#1}{#2}}

\newbox{\ORCIDicon}
\sbox{\ORCIDicon}{\large\includegraphics[width=0.8em]{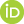}}

\begin{document}

\title{Functional renormalization group study of a four-fermion model with $CP$ violation
}

\author{Linlin \surname{Huang}\,\href{https://orcid.org/0009-0008-2174-8954}{\usebox{\ORCIDicon}}\,}
\email[]{huangll22@mails.jlu.edu.cn}
\affiliation{Center for Theoretical Physics and College of Physics, Jilin University, Changchun 130012, China}

\author{Mamiya \surname{Kawaguchi}\,\href{https://orcid.org/0000-0002-3103-1315}{\usebox{\ORCIDicon}}\,}
\email[]{mamiya@aust.edu.cn}
\affiliation{Anhui University of Science and Technology, Huainan, Anhui 232001, People’s Republic of China}

\author{Yadikaer \surname{Maitiniyazi}\,\href{https://orcid.org/0009-0004-0826-1130}{\usebox{\ORCIDicon}}\,}
\email[]{ydqem22@mails.jlu.edu.cn}
\affiliation{Center for Theoretical Physics and College of Physics, Jilin University, Changchun 130012, China}

\author{Shinya \surname{Matsuzaki}\,\href{https://orcid.org/0000-0003-4531-0363}{\usebox{\ORCIDicon}}\,}
\email[]{synya@jlu.edu.cn}
\affiliation{Center for Theoretical Physics and College of Physics, Jilin University, Changchun 130012, China}

 \author{Akio \surname{Tomiya}\,\href{https://orcid.org/0000-0001-9374-3716}{\usebox{\ORCIDicon}}\,}
 \email[]{akio@yukawa.kyoto-u.ac.jp}
 \affiliation{Department of Mathematics, Tokyo Woman’s Christian University, Tokyo 167-8585, Japan}
\affiliation{RIKEN Center for Computational Science, Kobe 650-0047, Japan} 

\author{Masatoshi \surname{Yamada}\,\href{https://orcid.org/0000-0002-1013-8631}{\usebox{\ORCIDicon}}\,}
\email[]{yamada@jlu.edu.cn}
\affiliation{Center for Theoretical Physics and College of Physics, Jilin University, Changchun 130012, China}

\begin{abstract}
We perform a functional renormalization group analysis of a four-fermion model with $CP$ and $P$ violation to explore the nonperturbative infrared dynamics of quantum chromodynamics (QCD) within the Wilsonian renormalization group framework, particularly in the context of spontaneous $CP$-violation models.  
Our analysis of the fixed-point structure reveals that, in the large-$N_c$ limit, the $CP$-violating $\bar{\theta}$ parameter is dynamically induced and approaches $\pi \cdot (N_f/2)$ (where $N_f$ is the number of flavors) as the system enters the chirally broken phase. This behavior arises due to criticality and the large anomalous dimensions of the $U(1)_A$-violating four-fermion couplings.  
Furthermore, this trend appears to persist beyond the leading large-$N_c$ approximation, provided that the infrared dynamics of QCD remains dominated by the scalar condensate of the quark bilinear, as expected. Notably, our findings highlight that $CP$-violating four-fermion interactions, which are perturbatively irrelevant, can become relevant in the chirally broken phase through nonperturbative effects, with potential implications for spontaneous $CP$-violation scenarios.
\end{abstract}
\maketitle

\section{Introduction}

The strong $CP$ problem—the question of why the parameter $\bar{\theta}$ is extremely small, with experimental constraints placing $\bar{\theta} < 10^{-11}$~\cite{Abel:2020pzs,Liang:2023jfj}—has been a major motivation for extending the Standard Model (SM). One possible resolution is the spontaneous $CP$-violation scenario~\cite{Nelson:1983zb,Barr:1984qx,Barr:1984fh,Babu:1989rb,Barr:1991qx,Mohapatra:1978fy,Beg:1978mt,Georgi:1978xz,Dine:1993qm,Dine:2015jga}, in which $CP$ and $P$ symmetries are imposed at the level of the classical action, ensuring that $\bar{\theta} = 0$ initially. A finite but small value of $\bar{\theta}$ then arises dynamically as a consequence of spontaneous $CP$ and $P$ symmetry breaking.

Loop corrections in this framework also generate effective $CP$- and $P$-violating operators, and their renormalization group (RG) evolution can induce a shift in $\bar{\theta}$. The perturbative RG evolution of loop-induced $CP$- and $P$-violating operators has been studied down to the intrinsic quantum chromodynamics (QCD) scale ($\sim 1$ GeV, comparable to the neutron mass scale) using the SM effective field theory approach~\cite{Jenkins:2017dyc,Hisano:2012cc,deVries:2021sxz,Banno:2023yrd}. Since these induced operators are typically of higher dimension, their RG running effects are expected to be small, ensuring that the shift in $\bar{\theta}$ remains modest.

However, certain induced higher-dimensional operators, including four-fermion interactions, play a crucial role in dynamical chiral symmetry breaking. Their effective couplings tend to diverge at a critical scale, signaling the onset of chiral symmetry breaking; see, e.g., Refs.~\cite{Aoki:1996fh,Aoki:2000dh,Aoki:2012mj}. In such strongly coupled regimes, perturbative approaches become inadequate, necessitating nonperturbative methods to accurately capture the RG evolution of $CP$-violating phases.

In this paper, we investigate the impact of four-fermion operators on the RG evolution of the $\bar\theta$ parameter using a simple four-fermion model with $CP$ (and $P$) violation. To track the RG flow of effective couplings, including $\bar\theta$, we employ the functional renormalization group (fRG) method~\cite{Wegner:1972ih,Wilson:1973jj,Polchinski:1983gv,Wetterich:1992yh}, which provides a Wilsonian RG formulation in quantum field theory.

Our model captures the essential features of the low-energy description around the dynamical chiral symmetry breaking scale. We analyze the fixed-point structure and demonstrate that, in the large-$N_c$ limit, a finite $\bar{\theta}$ is dynamically induced. Moreover, as the system approaches the chirally broken phase, $\bar{\theta}$ can reach the maximally $CP$-violating phase $\pi/2$ [or, more generally, $\bar{\theta} = \pi \cdot (N_f/2)$ for $N_f$ flavors with identical masses] due to criticality and the large anomalous dimensions of the $U(1)_A$-violating four-fermion couplings.
Note here that $U(1)_A$ is not to be confused with the Peccei-Quinn $U(1)$ symmetry.

This paper is organized as follows. In Sec.~\ref{sec:four-fermi model}, we first introduce the four-fermion model with $CP$ violation as a low-energy description of QCD, incorporating additional $CP$- and $P$-violating interactions beyond the strong $CP$ phase. 
Then we outline the fRG formalism used to study the RG flow of the $CP$-violating parameters. 
We present the fixed-point analysis of the system, demonstrating how $CP$ violation emerges dynamically due to the nonperturbative effects of chiral symmetry breaking. 
We analyze the RG flow of parameters and illustrate the behavior of the $CP$-violating phase $\bar\theta$ as the system evolves toward the infrared (IR) scale.
Finally, in Sec.~\ref{sec:summary}, we summarize our findings and discuss their implications for $CP$ violation and beyond the SM physics. 
Additional technical details, including derivations of the flow equations, are provided in Appendixes~\ref{app: Eulidean spacetime} and \ref{app:Derivation of flow equations}.

\section{Four-fermion model with $CP$ violation}
\label{sec:four-fermi model}

Motivated by the spontaneous $CP$-violation scenario, we consider a low-energy description of QCD with additional $CP$ and $P$ violation beyond the strong $CP$ phase to explore nonperturbative contributions to $CP$ and $P$ violation at scales $ \lesssim 1 $ GeV.  

To this end, in the present study, we employ the following one-flavor ($ N_f = 1 $) four-fermion theory as a toy model:  
\begin{align}
&S = \int d^4x \Bigg[ \bar{\psi}\left( i\gamma^\mu \partial_\mu - m e^{i \theta \gamma_5 / 2}\right) \psi \nn
&+ \frac{G_S}{2} \left(\bar{\psi} \psi\right)^2 
+ \frac{G_P}{2} \left(\bar{\psi} i \gamma_5 \psi\right)^2
+ C_4 \, (\bar{\psi}\psi ) (\bar{\psi} i \gamma_5 \psi) \Bigg].
\label{eq:toy model}
\end{align}
Here, $ \psi $ and $ \bar\psi $ are fermion fields that transform under the $ SU(N_c) $ color symmetry.  
The first two four-fermion operators in Eq.~\eqref{eq:toy model} (associated with $ G_S $ and $ G_P $, respectively) are invariant under the $P$ transformation, whereas the last term (associated with $ C_4 $) explicitly breaks $P$ invariance and, consequently, $CP$ invariance.\footnote{  
The transformation laws under the discrete symmetries $P$ and $C$ are given by  
\begin{align}
&\bar\psi \psi \sr{\text{P}}\to \bar\psi \psi,&
&\bar\psi i\gamma^5\psi \sr{\text{P}}\to -\bar\psi i\gamma^5\psi ,\nn
&\bar\psi \psi \sr{\text{C}}\to \bar\psi \psi,&
&\bar\psi i\gamma^5\psi  \sr{\text{C}}\to  \bar\psi i\gamma^5\psi .\nonumber
\end{align}  
}  
Thus, when $ C_4 = \theta = 0 $, the theory is CP-invariant.  
In the special case where $ G_S = G_P $ and $ C_4 = 0 $, the $ U(1)_A $ symmetry is restored, rendering $ \theta $ unphysical, as it can be eliminated by a $ U(1)_A $ rotation of the fermion fields.  

When extended to multiple flavors (e.g., two or three), the global non-Abelian chiral symmetry manifests through the four-fermion interactions in Eq.~\eqref{eq:toy model}, distinct from the explicitly broken Abelian $ U(1)_A $ symmetry (which results from $ G_S - G_P \neq 0 $ and $ C_4 \neq 0 $). For instance, in the two-flavor case, the $ G_S $ and $ G_P $ terms can be generalized as  
\begin{equation}
G_S \left[ (\bar{\psi}\psi)^2 + (\bar{\psi} i \gamma_5 \tau^a \psi)^2\right] 
+ G_P \left[ (\bar{\psi} \tau^a \psi)^2 + (\bar{\psi} i \gamma_5 \psi)^2\right],
\end{equation}  
where $ \psi $ is an $ SU(2) $-doublet fermion field, and $ \tau^a $ ($ a = 1,2,3 $) are the Pauli matrices. Each of the $ G_S $ and $ G_P $ terms individually preserves chiral $ SU(2) $ symmetry but breaks $ U(1)_A $ symmetry when $ G_S \neq G_P $.  

In this sense, the discrepancy between $ G_S $ and $ G_P $ effectively encodes the $ U(1)_A $ anomaly of the underlying QCD. The dynamical breaking of chiral $ SU(2) $ symmetry can be triggered solely by strong $ G_S $ and/or $ G_P $ couplings, similar to the Nambu--Jona-Lasinio mechanism, and does not necessarily require the presence of the $ U(1)_A $ anomaly ($ G_S \neq G_P $).  
Therefore, it is reasonable to conclude that dynamical chiral symmetry breaking generally occurs when $ G_S $ and/or $ G_P $ reach criticality, irrespective of the $ U(1)_A $ anomaly condition $ G_S \neq G_P $.

\subsection{Functional renormalization group}
We investigate the nonperturbative features arising from the four-fermion model in Eq.~\eqref{eq:toy model}, based on the 
fRG, a formulation of the Wilsonian renormalization group in quantum field theory.
The Wilsonian coarse-graining process is described by the functional differential equation for an effective action. 
In this work, we utilize the Wetterich equation~\cite{Wetterich:1992yh} in which the central object is the cutoff-dependent one-particle irreducible effective action $\Gamma_k$ where $k$ is the IR cutoff scale. 
See Refs.~\cite{Morris:1993qb,Reuter:1993kw,Ellwanger:1993mw,Morris:1998da,Berges:2000ew,Aoki:2000wm,Bagnuls:2000ae,Polonyi:2001se,Pawlowski:2005xe,Gies:2006wv,Delamotte:2007pf,Sonoda:2007av,Igarashi:2009tj,Rosten:2010vm} for a review on the fRG. 
The explicit form of the Wetterich equation (or flow equation) for a fermionic theory is given by
\begin{align}
\label{eq: Wetterich equation}
\p_t\Gamma_k =  -{\rm Tr} \left[ \left(\Gamma_k^{(2)}(p)+\mathcal R_k(p) \right)^{-1} \p_t \mathcal R_k(p) \right]\,,
\end{align}
where $\p_t=k\p_k$ is the derivative with respect to the dimensionless scale $t=\log (k/\Lambda)$ with a UV cutoff scale $\Lambda$. 
For this convention, the IR limit ($k\to 0$) corresponds to $t\to -\infty$.
In Eq.~\eqref{eq: Wetterich equation}, 
\begin{align}
\Gamma_k^{(2)}(p)=\frac{\overrightarrow\delta}{\delta\bar\psi(p)}\Gamma_k \frac{\overleftarrow\delta}{\delta\psi(p)}
\end{align}
is the full two-point function (inverse full propagator) for the fermion fields $\psi$ and $\bar\psi$ in momentum space, and ``Tr" denotes the trace acting on all spaces, e.g., momentum, color, flavor etc., on which the fields $\psi$ and $\bar\psi$ are defined. $\mathcal R_k(p)$ is the regulator function realizing the Wilsonian coarse-graining process in the path-integral formalism.
In this work, we employ the Litim-type cutoff function~\cite{Litim:2001up} for the fermion fields, i.e.,
\begin{align}
\mathcal R_k(p) = i{\Slash p}\left( \sqrt{\frac{k^2}{p^2}} -1 \right) \Theta(k^2-p^2)\,,
\label{eq:Litim cutoff}
\end{align}
where $\Theta(x)$ is the step function: $\Theta(x)=1$ for $x>0$, while $\Theta(x)=0$ for $x<0$.
For detailed discussions on the treatment of fermionic theories in the fRG framework, see Refs.~\cite{Salmhofer:2001tr,Kopietz:2010zz,Braun:2011pp,Metzner:2011cw}.

By solving the Wetterich equation~\eqref{eq: Wetterich equation} with the initial condition $\Gamma_\Lambda=S$ in Eq.~\eqref{eq:toy model} at $k=\Lambda$, we obtain the full effective action $\Gamma=\Gamma_{k=0}$ at $k=0$.
However, in the general Wilsonian viewpoint, $\Gamma_k$ is expressed as an infinite series of effective operators, making it infeasible to handle the effective action without approximations.
In this work, we make the following ansatz for the effective action for the model~\eqref{eq:toy model} in Euclidean spacetimes:
\begin{align}
&\Gamma_k \simeq \int \df^4x \Bigg[ \bar{\psi}\left( \gamma_\mu \partial_\mu + m e^{i \theta \gamma_5 / 2}\right) \psi \nn
&
-  \frac{G_S}{2}\left(\bar{\psi} \psi\right)^2 
-\frac{G_P}{2}\left(\bar{\psi} i \gamma_5 \psi\right)^2
-C_4 \, (\bar{\psi}\psi ) (\bar{\psi} i \gamma_5 \psi) \Bigg].
\label{eq: effective action for NJL}
\end{align}
Our convention for the Euclidean signature is summarized in Appendix~\ref{app: Eulidean spacetime}.
The couplings ($m$, $\theta$, $G_S$, $G_P$ and $C_4$) are $k$-dependent parameters. 
Four-fermion operators such as $(\bar\psi \gamma_\mu\psi)^2$, $(\bar\psi \gamma_\mu \gamma^5\psi)^2$, and $(\bar\psi \sigma_{\mu\nu}\psi)^2$ may give potentially the same order contributions as those four-fermion operators in the effective action~\eqref{eq: effective action for NJL}. Thus, working on the Fierz-complete basis is more general;  
see Refs.~\cite{Aoki:2009zza,Braun:2017srn,Braun:2018bik,Braun:2019aow}.
However, these operators generally remain irrelevant even at a nontrivial fixed point~\cite{Aoki:1999dv,Braun:2017srn}, so we neglect them in the present analysis.\footnote{
Earlier works~\cite{Aoki:1999dv,Braun:2017srn} have investigated the Fierz-complete basis in the effective action, given by  
\begin{align}
&\Gamma_k \simeq \int \df^4x \Bigg[ \bar{\psi} \gamma_\mu \partial_\mu \psi  
-  \frac{G_S}{2} \left( (\bar{\psi} \psi)^2 + (\bar{\psi} i \gamma_5 \psi)^2 \right) \nn  
&\qquad  
-\frac{G_V}{2} \left( (\bar{\psi} \gamma_\mu \psi)^2 + (\bar{\psi} i \gamma_\mu \gamma_5 \psi)^2 \right)  
\Bigg].
\end{align}  
It has been found that a nontrivial fixed point exists for $ G_S $ and $ G_V $, where the RG flows exhibit one IR-attractive and one IR-repulsive direction. This implies that, if the initial condition is set with $ G_S \neq 0 $ and $ G_V = 0 $, the vector-type coupling $ G_V $ is not a free parameter but rather an induced coupling. As a result, $ G_V $ remains irrelevant for the IR dynamics. This conclusion is expected to hold even in $CP$/$P$-violating systems, as $CP$ and $P$ symmetries do not influence the irrelevance of $ G_V $.  
}

Higher-dimensional operators, such as $(\bar\psi\psi)^3$, generally remain irrelevant in the IR regime even when nonperturbative effects are taken into account.
This is due to their large negative canonical dimensions, which make it unlikely for anomalous dimensions to be sizable enough to render these operators relevant.
Therefore, those would not substantially affect the dynamical chiral symmetry breaking.\footnote{
As we will see, the four-fermion coupling $ G_S $, which has a canonical dimension of $-2$, becomes relevant at a nontrivial fixed point, acquiring a scaling dimension of $+2$. This implies that the induced anomalous dimension is $+4$, which remains smaller than the canonical dimension of the coupling for $(\bar\psi\psi)^3$.  
In the bosonization framework, higher-dimensional fermionic operators correspond to higher-order terms in the bosonized potential, such as cubic and quartic interactions. These terms, however, do not significantly affect the curvature of the potential near the origin, meaning that they do not play a crucial role in triggering chiral symmetry breaking.
} 
In the current fermionic theory, the kinetic term does not receive quantum effects from the four-fermion operators thanks to the one-loop structure of the flow equation~\eqref{eq: Wetterich equation}.

\subsection{Flow equations}
Applying the Wetterich equation~\eqref{eq: Wetterich equation} to the effective action~\eqref{eq: effective action for NJL}, we derive the flow equations for the five couplings.
We list the flow equations for couplings in the effective action~\eqref{eq: effective action for NJL} within the large-$N_c$ approximation. 
The detailed derivation is presented in Appendix~\ref{app:Derivation of flow equations}.
Here, in order to discuss the fixed-point structure and the flow diagrams in the model in the following subsections, we define the dimensionless couplings as
\begin{align}
&\tilde m= mk^{-1},&
&\tilde G_S = G_Sk^2,\nn
&\tilde G_P = G_Pk^2,&
&\tilde C_4 = C_4k^2,
\end{align}
while $\theta$ is already dimensionless.

Although the fRG formalism allows us to take into account the large-$N_c$ subleading effects, those seem to be suppressed enough when the model is referred to as the underlying QCD theory.  This is because those do not dominantly participate in the dynamical chiral symmetry breaking via the scalar condensate of the quark bilinear, which governs the IR dynamics of QCD~\cite{Aoki:2014ola}. 
Thus, we study the flow equations within the large-$N_c$ approximation, in which the flow equations are obtained as
\begin{widetext}
\begin{align}
\label{eq:flow eq mass}
\p_t \tilde m
&= -\tilde m -4N_c\left[  (\tilde G_S +\tilde G_P)   +2\tilde C_4\sin\theta  + (\tilde G_S-\tilde G_P)\cos\theta\right] \tilde m\widetilde{\mathcal M}\,,\\[2ex]
\label{eq:flow eq theta}
\p_t \theta 
&= 8\left[ (\tilde G_S-\tilde G_P) \sin\theta
 - 2\tilde C_4 \cos\theta \right] \widetilde{\mathcal M}\,,\\[2ex]
\p_t \tilde G_S&= 2\tilde G_S -8
N_c\Big[
(\tilde G_S^2+\tilde C_4^2)(1 -\tilde m^2)
+ 2\tilde m^2( \tilde G_S^2 -\tilde C_4^2)\cos\theta 
+4\tilde m^2\tilde G_S\tilde C_4 \sin\theta
\Big]\widetilde{\mathcal I}\,,
\\[2ex]
\p_t \tilde G_P&= 2\tilde G_P - 8
N_c \Big[
(\tilde G_P^2+\tilde C_4^2)(1 -\tilde m^2)
- 2\tilde m^2( \tilde G_P^2 -\tilde C_4^2)\cos\theta 
+ 4\tilde m^2\tilde G_P \tilde C_4 \sin\theta
\Big]\widetilde{\mathcal I}\,,
\\[2ex]
\label{eq:flow eq C4}
\p_t \tilde C_4&= 2\tilde C_4 -8N_c
\Big[(\tilde G_S +\tilde G_P)\tilde C_4(1 -\tilde m^2)
+ 2\tilde m^2(\tilde G_S -\tilde G_P)\tilde C_4\cos\theta 
+ 2\tilde m^2 (\tilde G_S \tilde G_P+\tilde C_4^2) \sin\theta
\Big]\widetilde{\mathcal I}\,.
\end{align}
\end{widetext}
Here we have defined the (dimensionless) threshold functions as
\begin{align}
\widetilde{\mathcal M} &= \frac{1}{k^2}\int\frac{\df^4 p}{(2\pi)^4}\frac{p^2(1+r_k^f)\p_t r^f_k(p/k)}{(p^2(1+r^f_k)^2 + m^2)^2}\nn
&=\frac{1}{2(4\pi)^2}\frac{1}{(1 +\tilde m^2)^2} \,,\\[2ex]
\widetilde{\mathcal I} &=\int \frac{\df^4p}{(2\pi)^4}\frac{p^2(1+r_k^f) \p_t r^f_k(p/k)}{(p^2(1+r^f_k)^2 + m^2)^3}\nn
&= \frac{1}{2(4\pi)^2}\frac{1}{(1+\tilde m^2)^3}\,,
\end{align}
where in the second equality we have used the Litim-type cutoff function in Eq.~\eqref{eq:Litim cutoff}.

The $\beta$ function for the mass parameter is proportional to the mass parameter itself, which reflects the concept of technical naturalness~\cite{tHooft:1979rat,Wells:2013tta}. In the case where both $P$ and $U(1)_A$ symmetries are preserved, i.e., when $\tilde{G}_S = \tilde{G}_P$ and $\tilde{C}_4 = 0$, the $\beta$ function for $\theta$ vanishes, making $\theta$ invariant under RG transformations. Since the couplings $\tilde{G}_S$ and $\tilde{G}_P$ are $P$-invariant, their $\beta$ functions can depend on $G_S^2$ and $G_P^2$. In contrast, $\tilde{C}_4$ breaks the $P$ symmetry, so its $\beta$ function must vary under $P$ transformations and therefore cannot contain a $\tilde{C}_4^2$ term without another $P$ breaking effect, e.g., $\sin \theta \neq 0$.

\subsection{Fixed-point structure}
In this subsection, we study the fixed-point structure.
Thus, in the system, we solve $\p_t \tilde m = \p_t\theta = \p_t \tilde G_S=\p_t \tilde G_P=\p_t \tilde C_4 =0$ and find their solutions.
We start with the flow equation for $\theta$. The fixed-point condition $\p_t \theta=0$ yields a relation as
\begin{align}
\tan\theta^*= \frac{2\tilde C_4^*}{\tilde G_S^*-\tilde G_P^*}.
\label{eq:theta fixedpoint}
\end{align}
Inserting this into the other flow equations, the following nontrivial fixed points are found:
\begin{align}
\label{eq:FP1}
(\text{FP1}):&~~
\tilde m^*=0,\quad
\tilde G_S^*=\frac{8\pi^2}{N_c},\quad
\tilde G_P^*=0,\quad
\tilde C_4^*=0,\\
\label{eq:FP2}
(\text{FP2}):&~~
\tilde m^*=0,\quad
\tilde G_S^*=0,\quad
\tilde G_P^*=\frac{8\pi^2}{N_c},\quad
\tilde C_4^*=0,\\
\label{eq:FP3}
(\text{FP3}):&~~
\tilde m^*=0,\quad
\tilde G_S^*=G_P^*=\frac{8\pi^2}{N_c},\quad
\tilde C_4^*=0.
\end{align}
In addition to these nontrivial fixed points, there is the Gaussian (trivial) fixed point $\tilde m^*=\tilde G_S^*=\tilde G_P^*=\tilde C_4^*=0$.
FP1 and FP2 are $U(1)_A$-broken fixed points, while FP3 is $U(1)_A$-symmetric one.
For FP1 and FP2, from the fixed-point condition~\eqref{eq:theta fixedpoint}, $\tan\theta^*=0$, i.e. $\theta^*= n\pi$ with $n$ integers.
In the case of FP1, the right-hand side of Eq.~\eqref{eq:theta fixedpoint} becomes an indeterminate form. This fact is consistent with the unphysical nature of $\theta$ when the system becomes $U(1)_A$ invariant at FP3.

Next, we analyze the critical exponents (scaling dimensions). To this end, we linearize the flow equations. More specifically, for a set of the flow equations $\p_t \tilde g_i=\beta_i(\{ \tilde g\})$ where $\{\tilde g\}=\{\tilde m, \theta,\tilde G_S, \tilde G_P,\tilde C_4\}$, we perform a Taylor expansion around a fixed point $g_i^*$ and take up to the linear order of $\tilde g-\tilde g^*$ such that
\begin{align}
\p_t \tilde g_i\simeq \frac{\p\beta_i(\{ \tilde g\})}{\p \tilde g_j}\Bigg|_{\tilde g=\tilde g^*}(\tilde g_j -\tilde g_j^*)
\equiv -{\mathcal T}_{ij}(\tilde g_j -\tilde g_j^*),
\label{eq:linear flow equation}
\end{align}
where $\beta(\{g^*\})=0$ by definition of the fixed point.
The solution to Eq.~\eqref{eq:linear flow equation} is given by
\begin{align}
\tilde g_i = g_i^* + \sum_j c_j V_{i}^j \left(\frac{k}{\Lambda} \right)^{-\vartheta_j},
\end{align}
where $c_j$ are undetermined constants of integration, $\Lambda$ is a reference scale, and $V_i^j$ are the right eigenvectors of the stability matrix $\mathcal T$ with eigenvalues $\vartheta_i$ called the critical exponents or the scaling dimensions, namely, $\vartheta_i =-\text{eig}({\mathcal T})$.
When we approximate eigenvectors as $V_i^j\approx \delta_i^j$, the critical exponents are approximately obtained from the diagonal parts of $\mathcal T$. Since at all the fixed points we find $\tilde m^*=\tilde C_4^*=0$, such an approximation is reasonable and thus we have
\begin{align}
\vartheta_{m} &= -\frac{\p \beta_{m}}{\p \tilde m}\bigg|_{\tilde g=\tilde g^*}\nn
&= 1 - \frac{N_c}{8\pi^2}\left( (\tilde G_S^* + \tilde G_P^* ) - (\tilde G_S^* - \tilde G_P^* )\cos\theta^*\right),\\[2ex]
\vartheta_{C_4} &= -\frac{\p \beta_{C4}}{\p \tilde C_4}\bigg|_{\tilde g=\tilde g^*}
= -2 +\frac{N_c}{4\pi^2}\left(\tilde G_S^* + \tilde G_P^*\right),\\[2ex]
\vartheta_{\theta} &= -\frac{\p \beta_{\theta}}{\p \theta}\bigg|_{\tilde g=\tilde g^*}
=-\frac{N_c}{4\pi^2}\left( \tilde G_S^* -\tilde G_P^*\right)\cos\theta^*.
\end{align}

We exhibit the critical exponents at the Gaussian fixed point and FP3 for which the $CP$ phase $\theta$ is unphysical.
At the Gaussian fixed point, the critical exponents are identical with the canonical dimensions of couplings, i.e., we have $V_i^j=\delta_i^j$ and 
\begin{align}
&\vartheta_{m}=1,&
&\vartheta_{G_S} =-2,&
&\vartheta_{G_P} =-2,&
&\vartheta_{C_4} =-2,
\end{align}
while at FP3, we find
\begin{align}
&\vartheta_{m}=3,&
&\vartheta_{G_S} =2,&
&\vartheta_{G_P} =2,&
&\vartheta_{C_4} =2.
\end{align}
At FP3, all four-fermion couplings become relevant, indicating the emergence of a large anomalous dimension due to quantum effects that cannot be captured by perturbative methods. 
In particular, an important fact here is that $\tilde C_4$ becomes relevant at FP3 even if its fixed-point value is zero.
In the next subsection, we will see that $\tilde C_4$ is driven by $\tilde G_S$ and $\tilde G_P$ and grows in the IR regime.

At the nontrivial fixed points \eqref{eq:FP1} and \eqref{eq:FP2}, we compute eigenvalues and obtain, at FP1, 
\begin{align}
&\vartheta_m= 3,&
&\vartheta_\theta= -2,&
&\vartheta_{G_S} =2,&
&\vartheta_{G_P} =-2,&
&\vartheta_{C_4} =0,
\end{align}
and at FP2, 
\begin{align}
&\vartheta_m=1,&
&\vartheta_\theta=  2,&
&\vartheta_{G_S} =-2,&
&\vartheta_{G_P} =2,&
&\vartheta_{C_4} =0.
\end{align}
Since the $\beta$ functions for $\tilde m$ and $\theta$ depend on $\tilde G_S-\tilde G_P$, their critical exponents are also highly dependent on the differences of the fixed point values between $\tilde G_S^*$ and $\tilde G_P^*$.
As aforementioned, in QCD, the $U(1)_A$ symmetry is broken by the axial $U(1)$ anomaly and thus $\tilde G_S\neq \tilde G_P$. 
The condition $G_S > G_P$ is ensured by the positivity of the $\eta$ meson mass squared in a QCD hadron model with the two lightest quark flavors. This model features a chiral symmetry breaking and $U(1)_A$ anomaly structure similar to those discussed in Refs.~\cite{Frank:2003ve,Boer:2008ct,Boomsma:2009eh,Sakai:2011gs}.
In this sense, FP1~\eqref{eq:FP1} may be suitable as a low-energy effective field theory of QCD.

\begin{figure*}
\includegraphics[width=8cm]{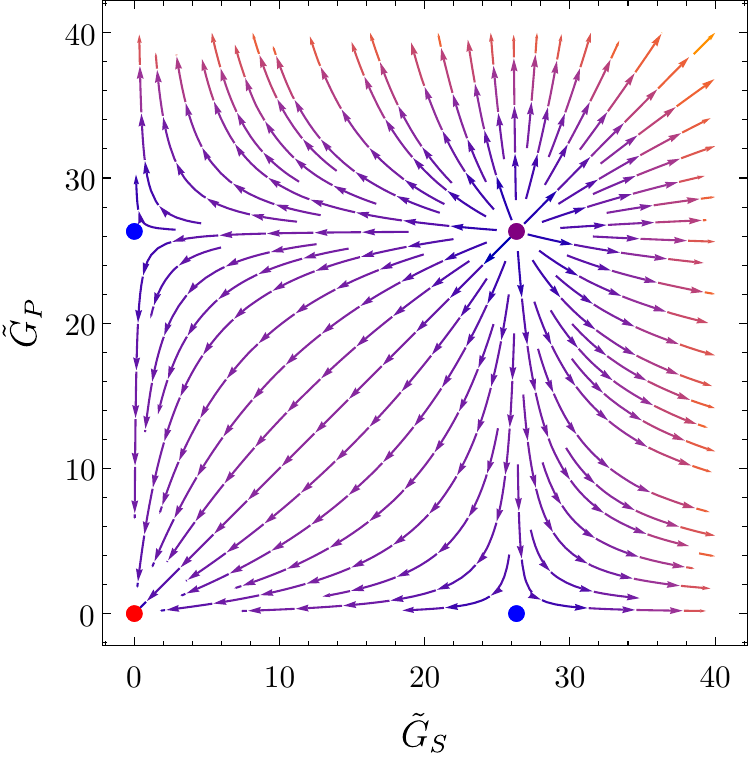}
\hspace{5ex}
\includegraphics[width=8cm]{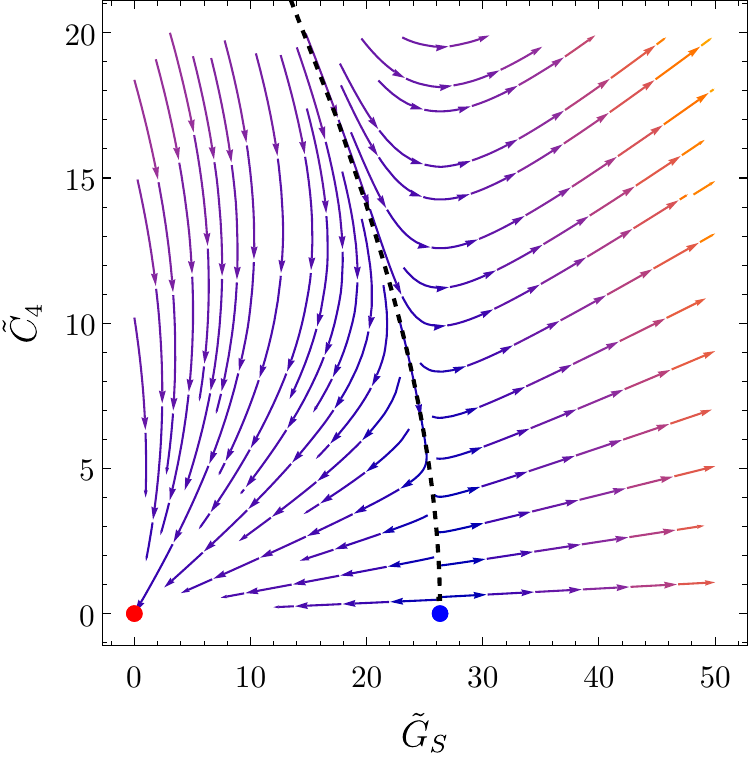}
\caption{
Flow diagrams ($N_c=3$) on $\tilde G_S$-$\tilde G_P$ plane with $\tilde C_4=0$ (left) and $\tilde G_S$-$\tilde C_4$ plane with $\tilde G_P=0$ (right) for $\tilde m=0$.
The red and purple points denote the Gaussian fixed point and FP3~\eqref{eq:FP3}, respectively, while blue points show FP1~\eqref{eq:FP1} and FP2~\eqref{eq:FP1}. The dashed lines correspond to the critical surface.
}
\label{fig:flow diagrams} 
\end{figure*}

\subsection{RG flow of parameters}
We present the behavior of RG flows as solutions to Eqs.~\eqref{eq:flow eq mass}--\eqref{eq:flow eq C4} with $N_c=3$.
To begin with, let us examine the case $\tilde m=0$, in which the flow equations for the four-fermion couplings form a closed system. 
In Fig.~\ref{fig:flow diagrams}, we display flow diagrams on the $\tilde G_S$-$\tilde G_P$ plane (with $\tilde C_4=0$) and on the $\tilde G_S$-$\tilde C_4$ plane (with $\tilde G_P=0$). The arrows indicate the direction of RG flows from the UV to the IR.
At the fixed point FP3, both $\tilde G_S$ and $\tilde G_P$ become relevant, meaning that they are free parameters in this model. For initial values of $\tilde G_S$ and $\tilde G_P$ that exceed the fixed-point values $G_S^*$ and $G_P^*$, the RG flows diverge in the IR regime, suggesting dynamical breaking of the discrete axial symmetry; see, e.g., Refs.~\cite{Aoki:1996fh,Aoki:2000dh,Aoki:2012mj}.
\begin{figure*}
\includegraphics[width=8cm]{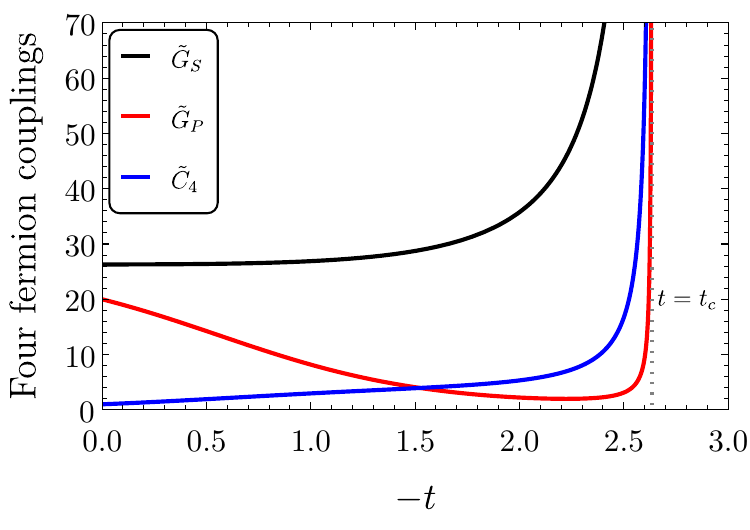}
\hspace{5ex}
\includegraphics[width=8cm]{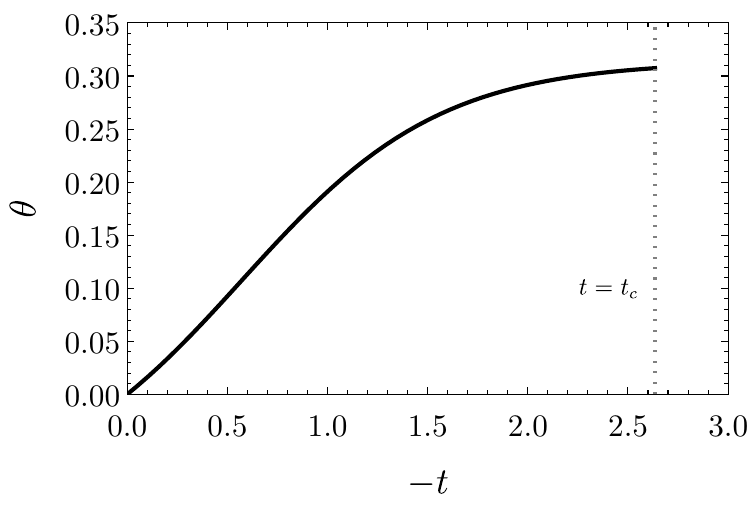}
\caption{
RG flow as a solution to Eqs.~\eqref{eq:flow eq mass}--\eqref{eq:flow eq C4} with $N_c=3$ and the initial condition $\tilde m=0.001$, $\theta=0$, $\tilde G_S=26.3$, $G_P=20$, and $\tilde C_4=1$ at $t=0$.
The vertical dotted line indicates the critical scale $t=t_c\simeq -2.64$.
}
\label{fig:flows} 
\end{figure*}
Interestingly, the right-hand panel of Fig.~\ref{fig:flow diagrams} shows that, in this broken phase, $\tilde C_4$ tends to grow in the IR direction.
This growth plays a crucial role in generating a finite $\theta$. Even if $\theta$ is initially set to zero at $t = 0$ ($k = \Lambda$), a nonzero $\tilde{C}_4$ can drive $\theta$ to finite values at IR scales.
In Fig.~\ref{fig:flows}, as an example, we present an RG flow solution to Eqs.~\eqref{eq:flow eq mass}--\eqref{eq:flow eq C4}, with the initial conditions $\tilde m=0.001$, $\theta=0$, $\tilde G_S=26.3$, $\tilde G_P=20$, and $\tilde C_4=1$ at $t=0$.
The RG flow of $\tilde m$ does not change from the initial value, so we do not show it.
The four-fermion couplings diverge at the critical scale $ t = t_c \simeq -2.64 $, corresponding to $ k_c = \Lambda e^{t_c} \simeq 0.071\Lambda $ in dimensionful terms. 
However, within our current framework, we are unable to track the RG flow beyond this critical scale. To do so, one would need to employ methods such as bosonization~\cite{Aoki:1999dw,Gies:2001nw,Gies:2002hq,Floerchinger:2009uf,Mitter:2014wpa,Braun:2014ata,Denz:2019ogb} or the weak RG approach~\cite{Aoki:2014ola,Aoki:2017rjl}, which lie beyond the scope of this work.

Here, we investigate the conditions that lead to maximal $P$/$CP$ violation ($\theta = \pi/2$) in the IR regime of our model.
To see this, we set $\tilde m=0$ and $\tilde G_P=0$ for simplicity.
Furthermore, we define $\tilde m_\sigma^2=1/\tilde G_S$ and $\rho=\tilde C_4/\tilde G_S$ whose flow equations read
\begin{align}
\label{eq:flow eq of msigma}
\p_t \tilde m_\sigma^2&= -2\tilde m_\sigma^2 + 8N_c (1 +\rho^2)\widetilde{\mathcal I},\\
\label{eq:flow eq of rho}
\p_t \rho &= 8N_c\frac{\rho^2}{\tilde m_\sigma^2}\widetilde{\mathcal I}.
\end{align}
Note here that the parameter $\tilde m_\sigma^2$ corresponds to the mass parameter of an emergent bosonic field (corresponding to the $\sigma$ meson) in the bosonization framework~\cite{Kodama:1999if,Aoki:2015hsa}. 
Thus, $\tilde m_\sigma^2=0$ corresponds to the critical value $\tilde G_S\to\infty$ at which the curvature of the effective potential for the bosonic field at the origin becomes zero.
When $\rho=0$, the RG flow of $\tilde m_\sigma^2$ can be followed beyond the critical scale without encountering singularities.
However, when $\rho$ is included, the RG flow of $\rho$ halts at $\tilde{m}_\sigma^2 = 0$ because its $\beta$ function is proportional to $1/\tilde{m}_\sigma^2$. 
As a result, the coupled system of \eqref{eq:flow eq of msigma} and \eqref{eq:flow eq of rho} develops a singularity, preventing us from extending the flow beyond the critical scale in this formulation. 
Therefore, we cannot go beyond the critical scale in the current setup.
Nevertheless, from a numerical perspective, the flow equation for the inverse four-fermion coupling~\eqref{eq:flow eq of msigma} remains a convenient choice for analysis.

\begin{figure*}
\includegraphics[width=8cm]{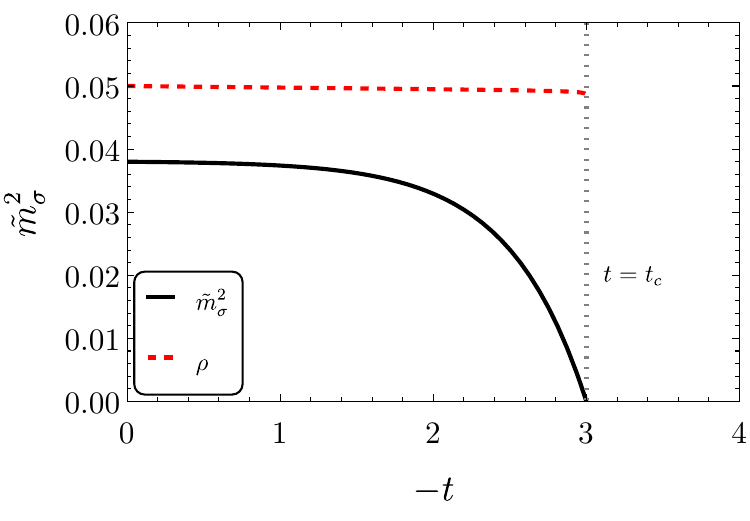}
\hspace{5ex}
\includegraphics[width=8cm]{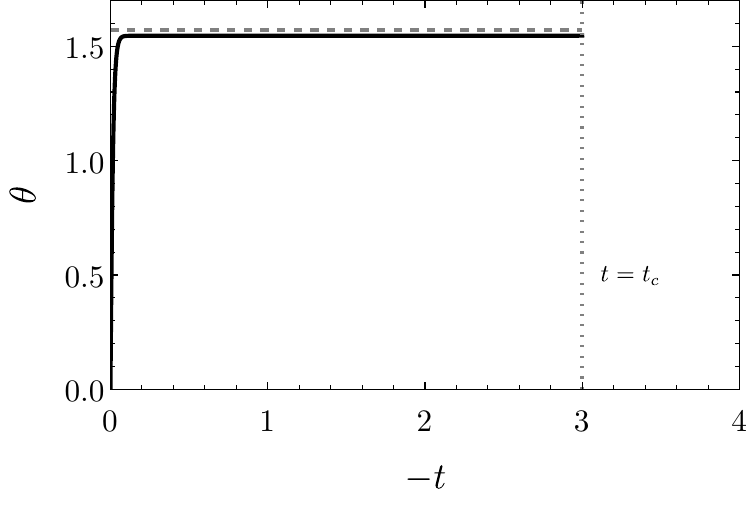}
\caption{
RG flow as a solution to Eqs.~\eqref{eq:flow eq of msigma} and \eqref{eq:flow eq of rho} with $N_c=3$ and the initial condition $\tilde m_\sigma^2 =3/8\pi^2$ and $\rho=0.05$ at $t=0$. The gray dashed line in the right-hand panel indicates the value of $\pi/2$. 
The vertical dotted line indicates the critical scale $t=t_c\simeq -3$.
}
\label{fig:flows 2} 
\end{figure*}

In the left-hand panel of Fig.~\ref{fig:flows 2}, we present the RG flow obtained by solving Eqs.~\eqref{eq:flow eq of msigma} and \eqref{eq:flow eq of rho} with the initial conditions $\tilde{m}_\sigma^2 = 3/8\pi^2$, $\rho = 0.05$, and $\theta = 0$ at $t = 0$. The RG flow of $\tilde{m}_\sigma^2$ reaches zero at the critical scale $t = t_c \simeq -3$.  
We note here that $\rho$ is a dimensionless parameter and exhibits a logarithmic running behavior, meaning it evolves gradually without significant variation under the RG flow.

From the right-hand panel of Fig.~\ref{fig:flows 2}, we observe that a finite value of $\theta$ emerges along with the solutions for $\tilde{m}_\sigma^2$ and $\rho$, eventually reaching $\pi/2$. Once $\theta$ attains $\pi/2$, the contribution of $\tilde{C}_4$ in the flow equation~\eqref{eq:flow eq theta} vanishes due to $\cos(\pi/2) = 0$.  
This mechanism is crucial in ensuring that $\theta$ does not exceed $\pi/2$, allowing us to analyze $P$ and $CP$ violation within the range $0 \leq \theta \leq \pi/2$ in the case of $N_f = 1$. In other words, the RG flow of $\theta$ does not exhibit limit-cycle behavior.

\section{Summary}
\label{sec:summary}

We have shown that the perturbatively irrelevant $CP$ (and $P$) violating four-fermion operators can be relevant due to the nonperturbatively yielded large anomalous dimensions triggered by 
the criticality of the dynamical chiral symmetry breaking. 
Consequently, the strong $CP$ violation can be amplified in the chiral broken phase with the presence of the $U(1)_A$ anomaly and the 
$CP$ phase approaches $\pi \cdot (N_f/2)$, even if it is tuned to vanish in the (seemingly) perturbative regime of QCD at the scales $\gtrsim 1$ GeV. 
Although this trend has been clarified in the large-$N_c$ limit by working on a four-fermion model, the $1/N_c$ subleading four-fermion term corrections seem to be suppressed enough.  
This is because those do not dominantly participate in the dynamical chiral symmetry breaking via the scalar condensate of the quark bilinear, 
which governs the IR dynamics of QCD. 
Thus, our conclusions remain robust.

The setup in this work allows us to investigate the RG flow of $\theta$ until the critical scale of dynamical chiral symmetry breaking. 
The further IR property of strong $CP$ violation is, however, of importance. 
The measurement of the neutron electron dipole moment is indeed physics with the photon transfer energy (the applied electric field strength) $\sim 10^{-6}$--$10^{-5}$ GeV~\cite{Abel:2020pzs,Liang:2023jfj}, which is much less than the typical QCD scale $\sim 0.1$--$1$ GeV. 
The state-of-the-art theoretical estimates based on the lattice QCD at the physical point~\cite{Alexandrou:2020mds} 
have been done for the transfer momentum in a range $\sim 200$--$400$ MeV, in which, however, the measurement is irrespective to $\bar{\theta}$, and hence does not address its transfer momentum dependence, i.e., the RG scale. 
To extend the RG flow analysis beyond the scale of dynamical chiral symmetry breaking within the fRG framework, additional techniques such as (dynamical) bosonization~\cite{Aoki:1999dw,Gies:2001nw,Gies:2002hq,Floerchinger:2009uf,Mitter:2014wpa,Braun:2014ata,Denz:2019ogb} and the weak formulation~\cite{Aoki:2014ola,Aoki:2017rjl} may be necessary.
We plan to explore these approaches in future studies on $CP$-violating four-fermion models.

Our findings have significant implications for modeling spontaneous $CP$-violation scenarios. To cancel the induced $\bar{\theta} = \pi \cdot (N_f/2)$, additional CP-violating contributions from physics beyond the QCD sector would be required at energy scales above the QCD intrinsic scale. Otherwise, the associated electric dipole moments (EDMs), including the quark EDM, chromo EDM, and neutron EDM, could be dynamically amplified.

\begin{acknowledgements}
The work of S.\,M. was supported in part by the National Science Foundation of China (NSFC) under Grant No.11747308, 11975108, 12047569 and the Seeds Funding of Jilin University. 
The work of M.\,K. is supported by RFIS-NSFC under Grant No.~W2433019.
The work of A.\,T. was partially supported by JSPS  KAKENHI Grant Numbers 20K14479, 22K03539, 22H05112, and 22H05111, and MEXT as ``Program for Promoting Researches on the Supercomputer Fugaku'' (Simulation for basic science: approaching the new quantum era, Grant Number JPMXP1020230411, and Search for physics beyond the Standard Model using large-scale lattice QCD simulation and development of AI technology toward next-generation lattice QCD, Grant Number JPMXP1020230409). 
The work of M.\,Y. is supported by the National Science Foundation of China (NSFC) under Grant No.~12205116 and the Seeds Funding of Jilin University.
\end{acknowledgements}

\onecolumngrid
\appendix
\section{Euclidean spacetime}
\label{app: Eulidean spacetime}
In this appendix, we provide our convention for the conversion from the Minkowski signature to the Euclidean one.

The world line in Minkowski spacetime is given by
\begin{align}
\df s^2=\df t^2-\df {\bvec x}^2=g_{\mu\nu}\df x^\mu \df x^\nu\,,
\end{align}
where the metric is defined as $g_{\mu\nu}=g^{\mu\nu}= {\rm diag}(1,-1,-1,-1)$.
Now, we introduce the coordinate vectors $x_{E\mu}$ in Euclidean spacetime as
\begin{align}
t=x^0&=-i x_{E0}=-i\tau\,,&
x^i &= x_{Ei}\,.
\end{align}
Then, the norm becomes
\begin{align}
x_\mu x^\mu
=(x^0)^2-(x^i)^2
=(-i x_{E0})^2-(x_{Ei})^2
=-(x_{E0}^2+x_{Ei}^2)
=-x_{E\mu}x_{E\mu}\,.
\end{align}
In Euclidean spacetime, we do not have to take care about covariance and contravariance of vectors.
Below, we summarize quantities in Euclidean spacetime
\begin{itemize}
\item[a)] Vector $V_\mu=\{ x_\mu, p_\mu, \cdots\}$,
\begin{align}
V^0&=-i V_{E0}\,,&
V^i &= V_{Ei}\,.
\label{appeq: vector 1 Euclidean}
\end{align}
Then we have
\begin{align}
V_\mu V^\mu=-V_{E\mu}V_{E\mu}\,.
\end{align}
\item[b)] Derivative operator $\p_\mu$,
\begin{align}
\p_\mu=\left( \frac{\p }{\p x^0},~ \frac{\p}{\p x^i}\right)
=\left( \frac{\p }{\p (-ix_{E0})},~ \frac{\p}{\p x_{Ei}}\right)
=\left( i\frac{\p }{\p x_{E0}},~ \frac{\p}{\p x_{Ei}}\right)
=(i\p_{E0},~\p_{Ei})\,.
\end{align}

\item[c)] Gamma matrices $\gamma_{\mu}$,
\begin{align}
\gamma^0&=\gamma_{E0}\,,&
\gamma^i &= i\gamma_{Ei}\,.
\label{appeq: gamma Eulidean}
\end{align}
Then, the Clifford algebra $\{ \gamma^\mu, \gamma^\nu\} =2g^{\mu\nu}$ in Minkowski spacetime becomes
\begin{align}
\{ \gamma^{0}, \gamma^{0}\}&=\{ \gamma_{E0}, \gamma_{E0}\}=2\,,&
\{ \gamma^{i}, \gamma^{j} \}&=\{ i\gamma_{Ei}, i\gamma_{Ej} \} 
=-\{ \gamma_{Ei}, \gamma_{Ej} \} =-2\delta_{ij}\,.
\end{align}
Thus, the Clifford algebra in Euclidean spacetime reads
\begin{align}
\{ \gamma_{E\mu},\gamma_{E\nu}\}=2\delta_{\mu\nu}
=2\pmat{
1 & & & \\
  & 1 & &\\
 & & 1 &\\
 & & & 1
}\,.
\end{align}

The hermiticity of $\gamma$ matrices is
\begin{align}
&(\gamma^0)^\dagger = \gamma^0 \Longrightarrow (\gamma_{E0})^\dagger = \gamma_{E0}\,,
&(\gamma^i)^\dagger = -\gamma^i \Longrightarrow (\gamma_{Ei})^\dagger = \gamma_{Ei}\,.
\end{align}
Thus, we have
\begin{align}
(\gamma_{E\mu})^\dagger = \gamma_{E\mu}\,.
\end{align}
The chirality matrix $\gamma^5$ is
\begin{align}
\gamma^5 = i\gamma^0\gamma^1\gamma^2\gamma^3 
= i\gamma_{E0}i\gamma_{E1}i\gamma_{E2}i\gamma_{E3}
= \gamma_{E0}\gamma_{E1}\gamma_{E2}\gamma_{E3}
= \gamma_{E5}\,.
\end{align}

\item[d)] Fields,
\begin{itemize}
\item Scalar,
\begin{align}
\phi\fn{x}=\phi_E\fn{x_E}\,.
\end{align}
\item Spinor,
\begin{align}
\psi\fn{x}&=\psi_E\fn{x_E},&
{\bar \psi}\fn{x}&=\psi^\dagger \gamma^0\fn{x}
=\psi^\dagger_E\gamma_{E0}\fn{x_E}
={\bar \psi}_E\fn{x_E}\,.
\end{align}
\item Vector,
\begin{align}
A^0\fn{x}&=-i A_{E0}\fn{x_E}\,,&
A^i\fn{x} &= A_{Ei}\fn{x_E}\,.
\label{appeq: vector Euclidean}
\end{align}
\end{itemize}

Note that $\bar\psi \gamma^\mu \psi$ is a vector; however, because of Eq.~\eqref{appeq: gamma Eulidean}, we have
\begin{align}
&\bar\psi \gamma^0 \psi =  \bar\psi_E \gamma_{E0} \psi_E\,,& 
&\bar\psi \gamma^i \psi =  i\bar\psi_E \gamma_{Ei} \psi_E\,,
\end{align}
which is different from Eqs.~\eqref{appeq: vector 1 Euclidean} and \eqref{appeq: vector Euclidean}.
The vector-type four-Fermion interaction becomes
\begin{align}
(\bar\psi \gamma^\mu \psi)^2
&=(\bar\psi \gamma^0 \psi)(\bar\psi \gamma^0 \psi) - (\bar\psi \gamma^i \psi)(\bar\psi \gamma^i \psi) \nn
&=(\bar\psi_E \gamma_{E0} \psi_E)(\bar\psi_E \gamma_{E0} \psi_E)  -(\bar\psi_E i\gamma_{Ei} \psi_E)(\bar\psi_E i\gamma_{Ei} \psi_E) \nn
&=(\bar\psi_E \gamma_{E0} \psi_E)(\bar\psi_E \gamma_{E0} \psi_E)  +(\bar\psi_E \gamma_{Ei} \psi_E)(\bar\psi_E \gamma_{Ei} \psi_E) \nn
&= (\bar\psi_E \gamma_{E\mu} \psi_E)^2\,.
\end{align}

\item[e)] Fourier transformations,
\al{
\phi_E\fn{x_E}&=\int \frac{\df ^4p_E}{(2\pi)^4}e^{ip_E\cdot x_E} {\tilde \phi}_E\fn{p_E}\,,\\
\psi_E\fn{x_E}&=\int \frac{\df ^4p_E}{(2\pi)^4}e^{ip_E\cdot x_E} {\tilde \psi}_E\fn{p_E}\,,\\
{\bar \psi}_E\fn{x}&=\int \frac{\df ^4p_E}{(2\pi)^4}e^{-ip_E\cdot x_E} {\bar {\tilde \psi}}_E\fn{p_E}\,.
}
\end{itemize}
Hereafter, we omit tildes on the Fourier modes.

In this convention for the Euclidean signature, the action for the spinor field reads
\begin{align}
S&=\int \df^4x \left[
{\bar \psi}\fn{x}i{\Slash \p}\psi\fn{x} - m{\bar \psi}\fn{x}\psi\fn{x} - V(\bar\psi,\psi)
\right]\nn
&=-i\int \df^4x_E \left[
-{\bar \psi}_E\fn{x_E}(\gamma_{E\mu}\p_{E\mu})\psi_E\fn{x_E}- m{\bar \psi}_E\fn{x_E}\psi_E\fn{x_E}  - V_E(\bar\psi_E,\psi_E)
\right]\nn
&=iS_E\,,
\end{align}
where
\begin{align}
S_E&=\int \df^4x_E \left[
{\bar \psi}_E\fn{x_E}(\Slash \p_{E})\psi_E\fn{x_E} + m{\bar \psi}_E\fn{x_E}\psi_E\fn{x_E}+ V_E(\bar\psi_E,\psi_E)
\right]\,.
\end{align}
In the main text and hereafter, we omit the subscript ``$E$" denoting the Euclidean signature.

\section{Derivation of flow equations}
\label{app:Derivation of flow equations}

The central method in this work is the Wetterich equation whose form is given by
\begin{align}
\p_t \Gamma_k = \frac{1}{2} {\rm STr}\left[ \left(\Gamma_k^{(2)}+\mathcal R_k \right)^{-1} \p_t \mathcal R_k \right] 
= \frac{1}{2} \tilde\p_t{\rm STr}\log \left[\Gamma_k^{(2)}+\mathcal R_k  \right]\,,
\end{align}
where $\tilde\p_t = (\p_t \mathcal R_k)\frac{\p}{\p \mathcal R_k}$ is the scale derivative acting only on the regulator function.
In general, the Hessian $\Gamma_k^{(2)}+\mathcal R_k$ is given as the supermatrix in superfield space. Namely, we can write in the form of
\begin{align}
\Gamma_k^{(2)}(p)+\mathcal R_k(p) 
=\frac{\overrightarrow\delta}{\delta\Phi^T(-p)}\Gamma_k\frac{\overleftarrow\delta}{\delta \Phi(p)} + \mathcal R_k(p)
=\begin{pmatrix}
M_{\rm BB} & M_{\rm BF} \\[2ex]
M_{\rm FB} & M_{\rm FF} 
\end{pmatrix}\,,
\end{align}
where $\Phi=(\phi,\psi,\bar\psi^T,\cdots)$ is the superfield.
In what follows, we consider the case of the fermionic system.

\subsection{Supermatrix, superdeterminant and supertrace}
We summarize the useful formulas for supermatrix to evaluate the Wetterich equation.
We begin by considering a supermatrix in a form of
\begin{align}\label{smatrix}
M=
\begin{pmatrix}
M_{\rm BB} & M_{\rm BF} \\[2ex]
M_{\rm FB} & M_{\rm FF} 
\end{pmatrix}\,.
\end{align}
Here, $M_{\rm BB}$ and $M_{\rm FF}$ are Grassmann-even elements, while $M_{\rm BF} $ and $M_{\rm FB}$ are Grassmann odd. We define the supertrace for the supermatrix~\eqref{smatrix} as
\begin{align}
{\rm str}~M={\rm tr}~M_{\rm BB}-{\rm tr}~M_{\rm FF}\,,
\label{eq: supertrace}
\end{align}
which satisfies
\begin{align}\label{defstr}
{\rm str}~(MN)={\rm str}~(NM)\,.
\end{align}
With this definition, the superdeterminant is defined by
\begin{align}\label{defsdet}
{\rm sdet}~M=\exp ({\rm str} ~{\rm ln}~M)\,,
\end{align}
such that
\begin{align}\label{sumatripro}
{\rm sdet}~(MN)={\rm sdet}~M\cdot {\rm sdet}~N\,.
\end{align}

Now, we deform $M$ as
\begin{align}
M=
\begin{pmatrix}
M_{\rm BB} & 0 \\[2ex]
M_{\rm FB} & 1
\end{pmatrix}
\begin{pmatrix}
1 & M_{\rm BB}^{-1}M_{\rm BF} \\[2ex]
0 & N_{\rm FF}
\end{pmatrix}  \,,
\label{eq: deformation 1}
\end{align}
where we have defined 
\begin{align}
N_{\rm FF}=M_{\rm FF}-M_{\rm FB}M_{\rm BB}^{-1}M_{\rm BF}\,.
\end{align}
From Eq.~\eqref{sumatripro}, we have
\begin{align}
{\rm sdet}~M=
{\rm sdet}
\begin{pmatrix}
M_{\rm BB} & 0 \\[2ex]
M_{\rm FB} & 1
\end{pmatrix}
{\rm sdet}
\begin{pmatrix}
1 & M_{\rm BB}^{-1}M_{\rm BF} \\[2ex]
0 & N_{\rm FF}
\end{pmatrix}\,.
\end{align}
Here, using Eq.~\eqref{defsdet} gives
\begin{align}
{\rm sdet}
\begin{pmatrix}
M_{\rm BB} & 0 \\[2ex]
M_{\rm FB} & 1
\end{pmatrix}
&=\exp \left\{ {\rm str}~{\rm ln}\left[
\begin{pmatrix}
1 & 0 \\[1ex]
0 & 1
\end{pmatrix}
+
\begin{pmatrix}\nonumber
M_{\rm BB}-1 & 0 \\[1ex]
M_{\rm FB} & 1
\end{pmatrix}
\right]
\right\} \\
\nonumber
&=\exp \left\{ {\rm str} \displaystyle \sum ^{\infty}_{n=1} \frac{(-)^{n-1}}{n}
\begin{pmatrix}
M_{\rm BB}-1 & 0 \\[1ex]
M_{\rm FB} & 1
\end{pmatrix}
\right\} \\
\nonumber
&=\exp \left\{ {\rm tr} \displaystyle \sum ^{\infty}_{n=1} \frac{(-)^{n-1}}{n}(M_{\rm BB}-1)^n
\right\} \\
\nonumber 
&=\exp \{ {\rm tr~ln}~M_{\rm BB}\} \\
&={\rm det}~M_{\rm BB}\,.
\end{align}
In the same manner, we obtain 
\begin{align}\nonumber
{\rm sdet}
\begin{pmatrix}
1 & M_{\rm BB}^{-1}M_{\rm BF} \\[2ex]
0 & N_{\rm FF}
\end{pmatrix}  
&=\exp \{ -{\rm tr~ln}~N_{\rm FF}\}
=({\rm det}~N_{\rm FF})^{-1}\,.
\end{align}
Note here that the minus sign arises from Eq.~\eqref{eq: supertrace}. Finally, we arrive at
\begin{align}
{\rm sdet}~M=\frac{{\rm det}~M_{\rm BB}}{{\rm det}~N_{\rm FF}}\,.
\end{align}
This implies that
\begin{align}
\log {\rm sdet}~M &= \log {\rm det}~M_{\rm BB} - \log {\rm det}(M_{\rm FF}-M_{\rm FB}M_{\rm BB}^{-1}M_{\rm BF})\nn
&= \tr \log M_{\rm BB} - \tr \log(M_{\rm FF}-M_{\rm FB}M_{\rm BB}^{-1}M_{\rm BF}) \,.
\end{align}

Instead of Eq.~\eqref{eq: deformation 1}, the supermatrix~\eqref{smatrix} can also be deformed as
\begin{align}
M=
\begin{pmatrix}
N_{\rm BB} & M_{\rm BF}M_{\rm FF}^{-1} \\[2ex]
0 & 1
\end{pmatrix}
\begin{pmatrix}
1 & 0\\[2ex]
M_{\rm FB} & M_{\rm FF}
\end{pmatrix}\,.
\end{align}
Here, we have defined
\begin{align}
N_{\rm BB}= M_{\rm BB}-M_{\rm BF}M_{\rm FF}^{-1}M_{\rm FB}
\end{align}
Then, the superdeterminant for $M$ can be written in the form as
\begin{align}
{\rm sdet}~M=\frac{{\rm det }~N_{\rm BB}}{{\rm det}~M_{\rm FF}}\,,
\end{align}
for which we have
\begin{align}
\log {\rm sdet}~M &= \log {\rm det}~(M_{\rm BB}-M_{\rm BF}M_{\rm FF}^{-1}M_{\rm FB}) - \log {\rm det} M_{\rm FF}\nn
&= \tr \log (M_{\rm BB}-M_{\rm BF}M_{\rm FF}^{-1}M_{\rm FB}) - \tr \log M_{\rm FF} \,.
\end{align}

\subsection{General structure of flow equation}
For the pure fermionic system, the supermatrix \eqref{smatrix} is 
\begin{align}
M \equiv \mathcal G_k^{-1}(p) =\frac{\overrightarrow\delta}{\delta\Phi^T(-p)}\Gamma_k\frac{\overleftarrow\delta}{\delta \Phi(p)} + \mathcal R_k(p) 
= \begin{pmatrix}
0 & 0 \\
0 & M_{\rm FF} 
\end{pmatrix}\,.
\end{align}
Here, we have the fermionic part in the field basis $\Phi(p)=\pmat{\psi(p) \\[1ex]\bar\psi^T(-p)}$ and $\Phi^T(-p)=\pmat{\psi^T(-p) & \bar\psi(p)}$ as
\begin{align}
M_{\rm FF} = \pmat{\displaystyle
\frac{\overrightarrow\delta}{\delta \psi^T(-p)}\mathcal G_k^{-1}(p) \frac{\overleftarrow \delta}{\delta\psi(p)} && \displaystyle\frac{\overrightarrow\delta}{\delta \psi^T(-p)}\mathcal G_k^{-1}(p) \frac{\overleftarrow\delta}{\delta \bar\psi^T(-p)}\\[3ex]
\displaystyle\frac{\overrightarrow\delta}{\delta \bar\psi(p)}\mathcal G_k^{-1}(p)\frac{\overleftarrow \delta}{\delta  \psi(p)} && \displaystyle\frac{\overrightarrow\delta}{\delta \bar\psi(p)}\mathcal G_k^{-1}(p)\frac{\overleftarrow\delta}{\delta  \bar\psi^T(-p)}
}\,.
\end{align}

We briefly summarize several techniques to derive the $\beta$ functions. For a given effective action $\Gamma_k$, we compute the second-order functional derivative. Together with the regulator function, one schematically has 
\begin{align}
\mathcal G_k^{-1}(p)\Big|_\text{FF}
= (\mathcal K_k + \mathcal R_k) + \mathcal V_k
= \mathcal P_k^{-1} + \mathcal V_k\,,
\label{eq: inverse propagator}
\end{align}
where $\mathcal K_k$ is the field-independent part, while $\mathcal V_k$ includes the vertex terms depending on field variables. More specifically, for an effective action, we obtain
\begin{align}
\mathcal K_k(p,-p)
= \pmat{\displaystyle
\frac{\overrightarrow\delta}{\delta \psi^T(-p)}\Gamma_k \frac{\overleftarrow \delta}{\delta\psi(p)} && \displaystyle\frac{\overrightarrow\delta}{\delta \psi^T(-p)}\Gamma_k \frac{\overleftarrow\delta}{\delta \bar\psi^T(-p)}\\[3ex]
\displaystyle\frac{\overrightarrow\delta}{\delta \bar\psi(p)}\Gamma_k\frac{\overleftarrow \delta}{\delta  \psi(p)} && \displaystyle\frac{\overrightarrow\delta}{\delta \bar\psi^T(p)}\Gamma_k\frac{\overleftarrow\delta}{\delta  \bar\psi^T(-p)}
}\Bigg|_{\psi=\bar\psi=0}
=\pmat{
0 & K_k^T(p)\\
K_k(p) & 0
} (2\pi)^4\delta^4(0)
\,.
\end{align}
The regulator matrix is given by
\begin{align}
\mathcal R_k(p)
=\pmat{
0 & R^f_k(p)^T\\
R^f_k(p) & 0
}(2\pi)^4\delta^4(0)
= \pmat{
0 & i\Slash p^T r^f_k(p/k)\\
i\Slash p r^f_k(p/k) & 0
}(2\pi)^4\delta^4(0)\,.
\label{eq: regulator matrix}
\end{align}
There are many choices for the regulator $r^f_k(p/k)$. In this work, we employ the Litim-type cutoff~\cite{Litim:2001up} given as
\begin{align}
r^f_k(p/k) = \left( \sqrt{\frac{k^2}{p^2}} -1 \right) \Theta(k^2-p^2)\,.
\end{align}
Then, $\mathcal P_k$ is just the regulated inverse propagator such that 
\begin{align}
\mathcal P_k^{-1}(p,-p) = \pmat{
0 & (P_k^{-1})^T\\
P_k^{-1} & 0
}= \pmat{
0 & (K_k(p) + R^f_k(p))^T\\
K_k(p) + R^f_k(p) & 0
} (2\pi)^4\delta^4(0)\,.
\label{eq: kinetic term matrix}
\end{align}
For the vertex matrix, we write
\al{
\mathcal V_k = \pmat{
V_{11} & V_{12}\\[1ex]
V_{21} & V_{22}
}\,.
\label{eq: vertex matrix}
}
Note here that $V_{12}=-(V_{21})^T$.

Once we compute Eq.~\eqref{eq: inverse propagator}, the regulated full propagator is obtained by the expansion in terms of the vertex operator $\mathcal V_k$ such that
\begin{align}
\mathcal G_k(p)\Big|_\text{FF}=\left( \Gamma_k^{(2)} + \mathcal R_k \right)^{-1}
= \mathcal P_k - \mathcal P_k\mathcal V_k\mathcal P_k + \mathcal P_k\mathcal V_k\mathcal P_k\mathcal V_k\mathcal P_k + \cdots.
\end{align}
Inserting this into flow equation \eqref{eq: Wetterich equation} and using Eqs.~\eqref{eq: kinetic term matrix} and \eqref{eq: vertex matrix}, we have
\begin{align}
\p_t\Gamma_k &= \frac{1}{2}\tilde\p_t{\rm STr}\log \,\left[\mathcal G_k^{-1}|_\text{FF}\right]
=\frac{1}{2}\tilde\p_t{\rm STr}\log \,\left[\mathcal P_k^{-1} + \mathcal V_k\right]
\nn
&= \frac{1}{2}{\rm STr} \left[ \mathcal P_k  \p_t \mathcal R_k \right]
- \frac{1}{2}{\rm STr} \left[ \mathcal P_k\mathcal V_k\mathcal P_k \p_t \mathcal R_k \right]
+ \frac{1}{2}{\rm STr} \left[ \mathcal P_k\mathcal V_k\mathcal P_k\mathcal V_k\mathcal P_k \p_t \mathcal R_k \right]  +\cdots\nn
&= -\Tr\left[P_k\p_t R^f_k \right]
+\left( \frac{1}{2}\Tr \left[ P_k V_{21} P_k \p_t R_k^f \right]
+\frac{1}{2}\Tr \left[ P_k^T V_{12} P_k^T \p_t (R_k^f)^T \right] \right)\nn
&\qquad
- \Tr \left[ (P_k V_{21} P_k V_{21}\mathcal P_k - P_k V_{22}P_k^T V_{11}\mathcal P_k) \p_t R_k^f \right]
+\cdots\nn
&= -\Tr\left[P_k\p_t R^f_k \right]
+\Tr \left[ P_k V_{21} P_k \p_t R_k^f \right]
- \Tr \left[ (P_k V_{21} P_k V_{21} P_k - P_k V_{22}P_k^T V_{11} P_k) \p_t R_k^f \right]
+\cdots\,.
\label{eq: expansion of Wetterich equation}
\end{align}
Note that the minus sign in Eq.~\eqref{eq: expansion of Wetterich equation} reflects the definition of the supertrace~\eqref{eq: supertrace}. The first and second terms on the left-hand side of Eq.~\eqref{eq: expansion of Wetterich equation} correspond to the quantum corrections to the vacuum energy and the mass term. We are interested especially in the third term which gives the quantum corrections to the four-Fermi interactions.

\subsection{Application to four-Fermion model}
Here,  based on the previous formulas, we show the explicit computation for deriving the flow equations in the four-fermion model in Eq.~\eqref{eq: effective action for NJL}. The truncated effective action in our work is given by
\begin{align}
\Gamma_k = \int \df^4x \Bigg[ \bar{\psi}\left( \gamma_\mu \partial_\mu+m e^{i \theta \gamma_5 / 2}\right) \psi
-  \frac{G_S}{2}\left(\bar{\psi} \psi\right)^2 
-\frac{G_P}{2}\left(\bar{\psi} i \gamma_5 \psi\right)^2
-C_4 \, (\bar{\psi}\psi ) (\bar{\psi} i \gamma_5 \psi) 
\Bigg]\,.
\label{appeq: effective action for NJL}
\end{align}

In momentum space, the kinetic term is written in the form as
\begin{align}
\int \df^4x \, \bar{\psi}\left( \gamma_\mu \partial_\mu+m e^{i \theta \gamma_5 / 2}\right) \psi
&= \int\frac{\df^4p}{(2\pi)^4}\frac{\df^4q}{(2\pi)^4} \int \df^4x \, e^{i(p-q)\cdot x}\bar{\psi}(q)\left( i\gamma^\mu p_\mu+m e^{i \theta \gamma_5 / 2}\right) \psi(p)\nn
&=\int\frac{\df^4p}{(2\pi)^4}\bar{\psi}(p)\left( i\gamma_\mu p_\mu+m e^{i \theta \gamma_5 / 2}\right) \psi(p),
\end{align}
from which the kinetic matrix $\mathcal K_k$ in momentum space is found to be
\begin{align}
(\mathcal K_k(p,-p))_{IJ}= \pmat{
0 &&  \left( i\Slash{p} - me^{i \theta \gamma_5 / 2} \right)_{IJ}^T\\[2ex]
\left(i\Slash{p} + me^{i \theta \gamma_5 / 2} \right)_{IJ}  && 0
}(2\pi)^4\delta^4(0)\,.
\end{align}
Together with the regulator matrix~\eqref{eq: regulator matrix}, the regulated inverse propagator reads
\begin{align}
(\mathcal P_k^{-1}(p,-p))_{IJ} = \pmat{
0 &&  \left( i\Slash{p}(1+r^f_k) - me^{i \theta \gamma_5 / 2} \right)^T_{IJ}\\[2ex]
\left(i\Slash{p}(1+r^f_k) + me^{i \theta \gamma_5 / 2} \right)_{IJ}  && 0
}(2\pi)^4\delta^4(0)\,.
\label{eq: resugalted inverse propagator}
\end{align}
Its inverse form is 
\begin{align}
(\mathcal P_k(p,-p))_{IJ} = \pmat{
0 &&  \frac{\left( -i\Slash{p}(1+r^f_k) + me^{-i \theta \gamma_5 / 2}\right)_{IJ}}{p^2(1+r^f_k)^2 + m^2}\\[2ex]
\frac{ \left( -i\Slash{p}(1+r^f_k) - me^{-i \theta \gamma_5 / 2} \right)^T_{IJ} }{p^2(1+r^f_k)^2 + m^2}  && 0
}(2\pi)^4\delta^4(0)\,.
\end{align}

The vertex matrix \eqref{eq: vertex matrix} is calculated as
\begin{align}
\left(V_{11} \right)_{IJ}&= \frac{\overrightarrow\delta}{\delta \psi_{I}(-p)}\Gamma_k\frac{\overleftarrow \delta}{\delta \psi_{J}(p)}\nn
&=  \frac{G_S}{2} \Big[ \bar\psi_{i,M} \delta_{MI}\bar\psi_{N}\delta_{NJ}
 \Big]
 + \frac{G_P}{2} \Big[ \bar\psi_{M}(i\gamma_5)_{MI} \bar\psi_{N}(i\gamma_5)_{NJ} \Big] 
 +C_4 \Big[ \bar\psi_{i,M} \delta_{MI}\bar\psi_{N}(i\gamma_5)_{NJ}
 \Big]\,,
 \\[2ex]
\left(V_{22} \right)_{IJ}&= \frac{\overrightarrow\delta}{\delta \bar\psi_{I}(p)}\Gamma_k\frac{\overleftarrow \delta}{\delta \bar\psi_{J}(-p)} \nn
&=  \frac{G_S}{2} \Big[\delta_{IM} \psi_{M}\delta_{JN} \psi_{N}\Big] 
+ \frac{G_P}{2} \Big[ (i\gamma_5)_{IM} \psi_{M}  (i\gamma_5)_{JN}\psi_{N}\Big]
+ C_4  \Big[\delta_{IM} \psi_{M} (i\gamma_5)_{JN}\psi_{N} \Big] \,,
\\[2ex]
\left(V_{21} \right)_{IJ}&= \frac{\overrightarrow\delta}{\delta \bar\psi_{I}(p)}\Gamma_k\frac{\overleftarrow \delta}{\delta \psi_{J}(p)}\nn
&=- G_S \Big[\delta_{IJ}\bar\psi\psi\Big] 
-  G_P \Big[(i\gamma_5)_{IJ} \bar\psi i\gamma_5\psi\Big]
- C_4 \Big[\delta_{IJ}\bar\psi i\gamma_5 \psi\Big] 
- C_4\Big[ \bar\psi\psi(i\gamma_5)_{IJ}\Big]
\nn
&\quad
- G_S\Big[\delta_{IM} \psi_{M} \bar\psi_{N} \delta_{NJ}\Big]
- G_P\Big[  (i\gamma_5)_{IM} \psi_{M} \bar\psi_{N} (i\gamma_5)_{NJ}\Big]
- C_4\Big[\delta_{IM} \psi_{M}  \bar\psi_{N} (i\gamma_5)_{NJ} \Big]
- C_4\Big[(i\gamma_5)_{IM} \psi_{M}\bar\psi_{N} \delta_{NJ} \Big]
\,,\\[2ex]
\left(V_{12} \right)_{IJ}&= \frac{\overrightarrow\delta}{\delta \psi_{I}(-p)}\Gamma_k\frac{\overleftarrow \delta}{\delta \bar\psi_{J}(-p)} = -\left(V_{21} \right)_{IJ}\,.
\end{align}

The large-$N_c$ approximation corresponds to the following replacement: $(V_{11})_{IJ} \to 0$, $(V_{22})_{IJ} \to 0$,
\begin{align}
&(V_{21})_{IJ} \to (V_{21}^{\rm LN})_{IJ} =- G_S \Big[\delta_{IJ}\bar\psi\psi\Big] 
-  G_P \Big[(i\gamma_5)_{IJ} \bar\psi i\gamma_5\psi\Big]
- C_4\Big[\delta_{IJ}\bar\psi i\gamma_5 \psi\Big] 
- C_4\Big[ \bar\psi\psi(i\gamma_5)_{IJ}\Big],\\
&\left(V_{12} \right)_{IJ}\to -\left[(V_{21}^{\rm LN})\right]^T_{IJ}.
\end{align}

\subsubsection{Vacuum energy}
First, we evaluate the vacuum energy part which is independent of field operators. From the first term on the right-hand side of Eq.~\eqref{eq: expansion of Wetterich equation}, we obtain
\begin{align}
\p_t \rho_k=-\frac{1}{\Omega_4}\Tr\left[P_k\p_t R^f_k \right] &= -\int \frac{\df^4p}{(2\pi)^4}\tr\left[\frac{\left( -i\Slash{p}(1+r^f_k) + me^{-i \theta \gamma_5 / 2}\right)_{IJ}}{p^2(1+r^f_k)^2 + m^2} i(\Slash p)_{JK} \p_t r^f_k(p/k) \delta^\text{color}_{\alpha\beta}\right]\nn
&=-4N_c\int \frac{\df^4p}{(2\pi)^4}\left[\frac{p^2(1+r^f_k)}{p^2(1+r^f_k)^2 + m^2} \p_t r^f_k(p/k) \right]\nn
&=-\frac{4N_c}{(4\pi)^2}\int_0^{k^2} \df p^2\,p^2\left[\frac{k^2}{k^2 + m^2} \right]\,,
\end{align}
where $\Omega_4=\int\df^4x = (2\pi)^4\delta^4(0)$ is the four-dimensional spacetime volume and $\tr \delta^\text{color}_{\alpha\beta} =\delta^\text{color}_{\alpha\alpha}=N_c$.
This contribution, however, does not influence upon the spinor dynamics, so is irrelevant in the current work.

\subsubsection{Mass term}
From the second term on the right-hand side of Eq.~\eqref{eq: expansion of Wetterich equation}, we read off quantum corrections for the mass term with the CP and P violating angle $\theta$, such that
\begin{align}
&\bar\psi\p_t (m e^{i\theta \gamma_5/2}) \psi
= \p_t(m\cos(\theta/2))\bar\psi\psi + \p_t(m\sin(\theta/2))\bar\psi i\gamma^5 \psi
=\frac{1}{\Omega_4}\Tr \left[ P_k(V_{21}^{\rm LN}) P_k \p_t R_k^f \right]\nn
&
=\int\frac{\df^4 p}{(2\pi)^4}\tr \left[ \frac{\left( -i\Slash{p}(1+r^f_k) + me^{-i \theta \gamma_5 / 2}\right)_{IK}}{p^2(1+r^f_k)^2 + m^2} (V_{21}^{\rm LN})_{KL} \frac{\left( -i\Slash{p}(1+r^f_k) + me^{-i \theta \gamma_5 / 2}\right)_{LM}}{p^2(1+r^f_k)^2 + m^2} i(\Slash p)_{MJ} \p_t r^f_k(p/k) \right]\nn
&
=\int\frac{\df^4 p}{(2\pi)^4}\frac{\p_t r^f_k(p/k)}{(p^2(1+r^f_k)^2 + m^2)^2} \tr \left[ \left( -i\Slash{p}(1+r^f_k) + me^{-i \theta \gamma_5 / 2}\right)_{IK} (V_{21}^{\rm LN})_{KL}\left( -i\Slash{p}(1+r^f_k) + me^{-i \theta \gamma_5 / 2}\right)_{LM} i(\Slash p)_{MJ}  \right]\nn
&=
-\left[ 8N_c G_S  \cos\left(\frac{\theta}{2} \right) + 8N_cC_4 \sin\left(\frac{\theta}{2} \right)
\right]m \mathcal M(\bar\psi\psi) 
- \left[ 8N_cG_P  \sin\left(\frac{\theta}{2} \right)+ 8N_cC_4 \cos\left(\frac{\theta}{2} \right)
\right]m\mathcal M (\bar\psi i\gamma^5\psi)\,, 
\end{align}
where we have defined the threshold function as
\al{
\mathcal M = \int\frac{\df^4 p}{(2\pi)^4}\frac{p^2(1+r_k^f)\p_t r^f_k(p/k)}{(p^2(1+r^f_k)^2 + m^2)^2}
=\frac{1}{2(4\pi)^2}\frac{k^6}{(k^2 +m^2)^2} .
}
We can read off the flow equations as
\begin{align}
&\p_t (m \cos(\theta/2)) = - 8N_c G_S m \cos\left(\frac{\theta}{2} \right) \mathcal M
-  8N_c C_4 m \sin\left(\frac{\theta}{2} \right)\mathcal M \,,\\
&\p_t (m \sin(\theta/2)) =  - 8N_c G_Pm \sin\left(\frac{\theta}{2} \right) \mathcal M
- 8N_c C_4 m \cos\left(\frac{\theta}{2} \right)\mathcal M\,,
\end{align}
from which we have the flow equations for $m$ and $\theta$ as
\begin{align}
\p_t m
&=  \cos(\theta/2)\p_t (m \cos(\theta/2)) + \sin(\theta/2) \p_t (m \sin(\theta/2))\nn
&=- 4N_c\left[  (G_S +G_P)   + 2C_4\sin\theta  + (G_S-G_P)\cos\theta\right] m\mathcal M
\,,\\[2ex]
\p_t \theta &=  \frac{2}{m} \left[ -\sin(\theta/2)\p_t (m \cos(\theta/2))  + \cos(\theta/2)\p_t (m \sin(\theta/2)) \right]\nn
&= 8N_c\left[ ( G_S- G_P) \sin\theta
 -   2C_4 \cos\theta \right] {\mathcal M}\,.
\end{align}
Here we define the dimensionless couplings
\begin{align}
\label{appeq: dimensionless couplings}
&\tilde m= mk^{-1},&
&\tilde G_S = G_Sk^2,&
&\tilde G_P = G_Pk^2,&
&\tilde C_4 = C_4k^2.
\end{align}
Then, the flow equations for $\tilde m$ and $\theta$ read
\begin{align}
\label{appeq:flow eq mass}
\p_t \tilde m
&= -\tilde m -4N_c\left[  (\tilde G_S +\tilde G_P)   +2\tilde C_4\sin\theta  + (\tilde G_S-\tilde G_P)\cos\theta\right] \tilde m\widetilde{\mathcal M}\,,\\[2ex]
\label{appeq:flow eq theta}
\p_t \theta 
&= 8\left[ (\tilde G_S-\tilde G_P) \sin\theta
 - 2\tilde C_4 \cos\theta \right] \widetilde{\mathcal M}\,,
\end{align}
where the dimensionless threshold function is 
\begin{align}
\widetilde{\mathcal M} &= \frac{1}{k^2}\int\frac{\df^4 p}{(2\pi)^4}\frac{p^2(1+r_k^f)\p_t r^f_k(p/k)}{(p^2(1+r^f_k)^2 + m^2)^2}
=\frac{1}{2(4\pi)^2}\frac{1}{(1 +\tilde m^2)^2} \,.
\end{align}

\subsubsection{Four-Fermion interactions}
The third term on the right-hand side of Eq.~\eqref{eq: expansion of Wetterich equation} corresponds to quantum corrections for the four-fermion interactions which are obtained from the flow equation as 
\begin{align}
-\frac{\p_t G_S}{2} (\bar\psi\psi)^2 
-\frac{\p_t G_P}{2} (\bar\psi i\gamma^5\psi)^2
-\p_t C_4 (\bar\psi\psi)(\bar\psi i\gamma^5\psi)
= -\frac{1}{\Omega_4}\Tr \left[P_k V_{21}^\text{LN} P_k V_{21}^\text{LN}P_k  \p_t R_k^f \right].
\end{align}
We compute the right-hand side as
\begin{align}
& -\frac{1}{\Omega_4}\Tr \left[ P_k V_{21}^\text{LN} P_k V_{21}^\text{LN} P_k \p_t R_k^f \right]\nn
&= -\frac{1}{\Omega_4} \Tr \left[ \frac{ \left( -i\Slash{p}(1+r^f_k) + me^{-i \theta \gamma_5 / 2} \right) }{p^2(1+r^f_k)^2 + m^2}  V_{21} \frac{ \left( -i\Slash{p}(1+r^f_k) + me^{-i \theta \gamma_5 / 2} \right) }{p^2(1+r^f_k)^2 + m^2}  V_{21} \frac{ \left( -i\Slash{p}(1+r^f_k) + me^{-i \theta \gamma_5 / 2} \right) }{p^2(1+r^f_k)^2 + m^2} \p_t R_k^f \right]\nn
&= -\int \frac{\df^4p}{(2\pi)^4}\frac{p^2(1+r_k^f) \p_t r^f_k(p/k)}{(p^2(1+r^f_k)^2 + m^2)^3}
\Bigg[ 
-N_c\left\{
(4G_S^2+ 4C_4^2)(k^2 -m^2)
- 2m^2( 4G_S^2 -4C_4^2)\cos\theta -16m^2G_S C_4 \sin\theta
\right\} (\bar\psi\psi)^2\nn
&\qquad
-N_c \left\{
(4G_P^2+4C_4^2)(k^2 -m^2)
+ 2m^2( 4G_P^2 -4C_4^2)\cos\theta - 16m^2G_PC_4 \sin\theta
\right\} (\bar\psi i\gamma^5\psi)^2\nn
&\qquad
-4N_c
\left\{
2(G_S + G_P)C_4(k^2 -m^2)
- 4m^2( G_S -G_P)C_4\cos\theta - m^2 (4G_S G_P+4C_4^2) \sin\theta
\right\}(\bar\psi\psi) (\bar\psi i\gamma^5\psi)
\Bigg]\nn
&= 
\Bigg[ 
4N_c\left\{
(G_S^2+C_4^2)(k^2 -m^2)
+ 2m^2( G_S^2 -C_4^2)\cos\theta 
+4m^2G_SC_4 \sin\theta
\right\}\Bigg]\mathcal I (\bar\psi\psi)^2\nn
&\quad
+
\Bigg[
4N_c \left\{
(G_P^2+C_4^2)(k^2 -m^2)
- 2m^2( G_P^2 -C_4^2)\cos\theta 
+ 4m^2G_PC_4 \sin\theta
\right\} \Bigg]\mathcal I(\bar\psi i\gamma^5\psi)^2\nn
&\quad
+\Bigg[
8N_c
\left\{
(G_S + G_P)C_4(k^2 -m^2)
+ 2m^2( G_S -G_P)C_4\cos\theta 
+ 2m^2 (G_S G_P+C_4^2) \sin\theta
\right\} \Bigg]\mathcal I(\bar\psi\psi) (\bar\psi i\gamma^5\psi),
\end{align}
where we have defined the threshold function
\begin{align}
\mathcal I =\int \frac{\df^4p}{(2\pi)^4}\frac{p^2(1+r_k^f) \p_t r^f_k(p/k)}{(p^2(1+r^f_k)^2 + m^2)^3}
= \frac{1}{2(4\pi)^2}\frac{k^6}{(k^2+m^2)^3}.
\end{align}
Then, we obtain the flow equations
\begin{align}
-\frac{\p_t G_S}{2} &=4N_c\left\{
(G_S^2+C_4^2)(k^2 -m^2)
+ 2m^2( G_S^2 -C_4^2)\cos\theta 
+4m^2G_SC_4 \sin\theta
\right\}\mathcal I ,\\
-\frac{\p_t G_P}{2} &=4N_c \left\{
(G_P^2+C_4^2)(k^2 -m^2)
- 2m^2( G_P^2 -C_4^2)\cos\theta 
+ 4m^2G_PC_4 \sin\theta
\right\} \mathcal I ,\\
-\p_t C_4 &= 8N_c
\left\{
(G_S + G_P)C_4(k^2 -m^2)
+ 2m^2( G_S -G_P)C_4\cos\theta 
+ 2m^2 (G_S G_P+C_4^2) \sin\theta
\right\}\mathcal I.
\end{align}
For the dimensionless couplings~\eqref{appeq: dimensionless couplings}, we have
\begin{align}
\p_t \tilde G_S&= 2\tilde G_S -8
N_c\Big[
(\tilde G_S^2+\tilde C_4^2)(1 -\tilde m^2)
+ 2\tilde m^2( \tilde G_S^2 -\tilde C_4^2)\cos\theta 
+4\tilde m^2\tilde G_S\tilde C_4 \sin\theta
\Big]\widetilde{\mathcal I}\,,
\\[2ex]
\p_t \tilde G_P&= 2\tilde G_P - 8
N_c \Big[
(\tilde G_P^2+\tilde C_4^2)(1 -\tilde m^2)
- 2\tilde m^2( \tilde G_P^2 -\tilde C_4^2)\cos\theta 
+ 4\tilde m^2\tilde G_P \tilde C_4 \sin\theta
\Big]\widetilde{\mathcal I}\,,
\\[2ex]
\label{appeq:flow eq C4}
\p_t \tilde C_4&= 2\tilde C_4 -8N_c
\Big[(\tilde G_S +\tilde G_P)\tilde C_4(1 -\tilde m^2)
+ 2\tilde m^2(\tilde G_S -\tilde G_P)\tilde C_4\cos\theta 
+ 2\tilde m^2 (\tilde G_S \tilde G_P+\tilde C_4^2) \sin\theta
\Big]\widetilde{\mathcal I}\,,
\end{align}
with the dimensionless threshold function
\begin{align}
\widetilde{\mathcal I} &=\int \frac{\df^4p}{(2\pi)^4}\frac{p^2(1+r_k^f) \p_t r^f_k(p/k)}{(p^2(1+r^f_k)^2 + m^2)^3}
= \frac{1}{2(4\pi)^2}\frac{1}{(1+\tilde m^2)^3}\,.
\end{align}

\bibliographystyle{JHEP} 
\bibliography{refs}

\providecommand{\href}[2]{#2}\begingroup\raggedright\begin{thebibliography}{10}

\bibitem{Abel:2020pzs}
C.~Abel et~al., \emph{{Measurement of the Permanent Electric Dipole Moment of
  the Neutron}},
  \href{https://doi.org/10.1103/PhysRevLett.124.081803}{\emph{Phys. Rev. Lett.}
  {\bfseries 124} (2020) 081803}
  [\href{https://arxiv.org/abs/2001.11966}{{\ttfamily 2001.11966}}].

\bibitem{Liang:2023jfj}
{\scshape \ensuremath{\chi}QCD} collaboration, \emph{{Nucleon electric dipole
  moment from the \ensuremath{\theta} term with lattice chiral fermions}},
  \href{https://doi.org/10.1103/PhysRevD.108.094512}{\emph{Phys. Rev. D}
  {\bfseries 108} (2023) 094512}
  [\href{https://arxiv.org/abs/2301.04331}{{\ttfamily 2301.04331}}].

\bibitem{Nelson:1983zb}
A.~E. Nelson, \emph{{Naturally Weak CP Violation}},
  \href{https://doi.org/10.1016/0370-2693(84)92025-2}{\emph{Phys. Lett. B}
  {\bfseries 136} (1984) 387}.

\bibitem{Barr:1984qx}
S.~M. Barr, \emph{{Solving the Strong CP Problem Without the Peccei-Quinn
  Symmetry}}, \href{https://doi.org/10.1103/PhysRevLett.53.329}{\emph{Phys.
  Rev. Lett.} {\bfseries 53} (1984) 329}.

\bibitem{Barr:1984fh}
S.~M. Barr, \emph{{A Natural Class of Nonpeccei-quinn Models}},
  \href{https://doi.org/10.1103/PhysRevD.30.1805}{\emph{Phys. Rev. D}
  {\bfseries 30} (1984) 1805}.

\bibitem{Babu:1989rb}
K.~S. Babu and R.~N. Mohapatra, \emph{{A Solution to the Strong {CP} Problem
  Without an Axion}},
  \href{https://doi.org/10.1103/PhysRevD.41.1286}{\emph{Phys. Rev. D}
  {\bfseries 41} (1990) 1286}.

\bibitem{Barr:1991qx}
S.~M. Barr, D.~Chang and G.~Senjanovic, \emph{{Strong CP problem and parity}},
  \href{https://doi.org/10.1103/PhysRevLett.67.2765}{\emph{Phys. Rev. Lett.}
  {\bfseries 67} (1991) 2765}.

\bibitem{Mohapatra:1978fy}
R.~N. Mohapatra and G.~Senjanovic, \emph{{Natural Suppression of Strong p and t
  Noninvariance}},
  \href{https://doi.org/10.1016/0370-2693(78)90243-5}{\emph{Phys. Lett. B}
  {\bfseries 79} (1978) 283}.

\bibitem{Beg:1978mt}
M.~A.~B. Beg and H.~S. Tsao, \emph{{Strong P, T Noninvariances in a Superweak
  Theory}}, \href{https://doi.org/10.1103/PhysRevLett.41.278}{\emph{Phys. Rev.
  Lett.} {\bfseries 41} (1978) 278}.

\bibitem{Georgi:1978xz}
H.~Georgi, \emph{{A Model of Soft CP Violation}}, {\emph{Hadronic J.}
  {\bfseries 1} (1978) 155}.

\bibitem{Dine:1993qm}
M.~Dine, R.~G. Leigh and A.~Kagan, \emph{{Supersymmetry and the Nelson-Barr
  mechanism}}, \href{https://doi.org/10.1103/PhysRevD.48.2214}{\emph{Phys. Rev.
  D} {\bfseries 48} (1993) 2214}
  [\href{https://arxiv.org/abs/hep-ph/9303296}{{\ttfamily hep-ph/9303296}}].

\bibitem{Dine:2015jga}
M.~Dine and P.~Draper, \emph{{Challenges for the Nelson-Barr Mechanism}},
  \href{https://doi.org/10.1007/JHEP08(2015)132}{\emph{JHEP} {\bfseries 08}
  (2015) 132} [\href{https://arxiv.org/abs/1506.05433}{{\ttfamily
  1506.05433}}].

\bibitem{Jenkins:2017dyc}
E.~E. Jenkins, A.~V. Manohar and P.~Stoffer, \emph{{Low-Energy Effective Field
  Theory below the Electroweak Scale: Anomalous Dimensions}},
  \href{https://doi.org/10.1007/JHEP01(2018)084}{\emph{JHEP} {\bfseries 01}
  (2018) 084} [\href{https://arxiv.org/abs/1711.05270}{{\ttfamily
  1711.05270}}].

\bibitem{Hisano:2012cc}
J.~Hisano, K.~Tsumura and M.~J.~S. Yang, \emph{{QCD Corrections to Neutron
  Electric Dipole Moment from Dimension-six Four-Quark Operators}},
  \href{https://doi.org/10.1016/j.physletb.2012.06.038}{\emph{Phys. Lett. B}
  {\bfseries 713} (2012) 473}
  [\href{https://arxiv.org/abs/1205.2212}{{\ttfamily 1205.2212}}].

\bibitem{deVries:2021sxz}
J.~de~Vries, P.~Draper, K.~Fuyuto, J.~Kozaczuk and B.~Lillard,
  \emph{{Uncovering an axion mechanism with the EDM portfolio}},
  \href{https://doi.org/10.1103/PhysRevD.104.055039}{\emph{Phys. Rev. D}
  {\bfseries 104} (2021) 055039}
  [\href{https://arxiv.org/abs/2107.04046}{{\ttfamily 2107.04046}}].

\bibitem{Banno:2023yrd}
T.~Banno, J.~Hisano, T.~Kitahara and N.~Osamura, \emph{{Closer look at the
  matching condition for radiative QCD \ensuremath{\theta} parameter}},
  \href{https://doi.org/10.1007/JHEP02(2024)195}{\emph{JHEP} {\bfseries 02}
  (2024) 195} [\href{https://arxiv.org/abs/2311.07817}{{\ttfamily
  2311.07817}}].

\bibitem{Aoki:1996fh}
K.-I. Aoki, K.-i. Morikawa, J.-I. Sumi, H.~Terao and M.~Tomoyose,
  \emph{{Nonperturbative renormalization group analysis of the chiral critical
  behaviors in QED}}, \href{https://doi.org/10.1143/PTP.97.479}{\emph{Prog.
  Theor. Phys.} {\bfseries 97} (1997) 479}
  [\href{https://arxiv.org/abs/hep-ph/9612459}{{\ttfamily hep-ph/9612459}}].

\bibitem{Aoki:2000dh}
K.-I. Aoki, K.~Takagi, H.~Terao and M.~Tomoyose, \emph{{Nonladder extended
  renormalization group analysis of the dynamical chiral symmetry breaking}},
  \href{https://doi.org/10.1143/PTP.103.815}{\emph{Prog. Theor. Phys.}
  {\bfseries 103} (2000) 815}
  [\href{https://arxiv.org/abs/hep-th/0002038}{{\ttfamily hep-th/0002038}}].

\bibitem{Aoki:2012mj}
K.-I. Aoki and D.~Sato, \emph{{Solving the QCD non-perturbative flow equation
  as a partial differential equation and its application to the dynamical
  chiral symmetry breaking}},
  \href{https://doi.org/10.1093/ptep/ptt018}{\emph{PTEP} {\bfseries 2013}
  (2013) 043B04} [\href{https://arxiv.org/abs/1212.0063}{{\ttfamily
  1212.0063}}].

\bibitem{Wegner:1972ih}
F.~J. Wegner and A.~Houghton, \emph{{Renormalization group equation for
  critical phenomena}},
  \href{https://doi.org/10.1103/PhysRevA.8.401}{\emph{Phys. Rev. A} {\bfseries
  8} (1973) 401}.

\bibitem{Wilson:1973jj}
K.~G. Wilson and J.~B. Kogut, \emph{{The Renormalization group and the epsilon
  expansion}}, \href{https://doi.org/10.1016/0370-1573(74)90023-4}{\emph{Phys.
  Rept.} {\bfseries 12} (1974) 75}.

\bibitem{Polchinski:1983gv}
J.~Polchinski, \emph{{Renormalization and Effective Lagrangians}},
  \href{https://doi.org/10.1016/0550-3213(84)90287-6}{\emph{Nucl. Phys.}
  {\bfseries B231} (1984) 269}.

\bibitem{Wetterich:1992yh}
C.~Wetterich, \emph{{Exact evolution equation for the effective potential}},
  \href{https://doi.org/10.1016/0370-2693(93)90726-X}{\emph{Phys. Lett.}
  {\bfseries B301} (1993) 90}
  [\href{https://arxiv.org/abs/1710.05815}{{\ttfamily 1710.05815}}].

\bibitem{Morris:1993qb}
T.~R. Morris, \emph{{The Exact renormalization group and approximate
  solutions}}, \href{https://doi.org/10.1142/S0217751X94000972}{\emph{Int. J.
  Mod. Phys.} {\bfseries A9} (1994) 2411}
  [\href{https://arxiv.org/abs/hep-ph/9308265}{{\ttfamily hep-ph/9308265}}].

\bibitem{Reuter:1993kw}
M.~Reuter and C.~Wetterich, \emph{{Effective average action for gauge theories
  and exact evolution equations}},
  \href{https://doi.org/10.1016/0550-3213(94)90543-6}{\emph{Nucl. Phys.}
  {\bfseries B417} (1994) 181}.

\bibitem{Ellwanger:1993mw}
U.~Ellwanger, \emph{{Flow equations for N point functions and bound states,
  Proceedings, Workshop on Quantum field theoretical aspects of high energy
  physics: Bad Frankenhausen, Germany, September 20-24, 1993}},
  \href{https://doi.org/10.1007/BF01555911}{\emph{Z. Phys.} {\bfseries C62}
  (1994) 503} [\href{https://arxiv.org/abs/hep-ph/9308260}{{\ttfamily
  hep-ph/9308260}}].

\bibitem{Morris:1998da}
T.~R. Morris, \emph{{Elements of the continuous renormalization group,
  Nonperturbative QCD: Structure of the QCD vacuum: Proceedings, Yukawa
  International Seminar, YKIS'97, Kyoto, Japan, December 2-12, 1997}},
  \href{https://doi.org/10.1143/PTPS.131.395}{\emph{Prog. Theor. Phys. Suppl.}
  {\bfseries 131} (1998) 395}
  [\href{https://arxiv.org/abs/hep-th/9802039}{{\ttfamily hep-th/9802039}}].

\bibitem{Berges:2000ew}
J.~Berges, N.~Tetradis and C.~Wetterich, \emph{{Nonperturbative renormalization
  flow in quantum field theory and statistical physics}},
  \href{https://doi.org/10.1016/S0370-1573(01)00098-9}{\emph{Phys. Rept.}
  {\bfseries 363} (2002) 223}
  [\href{https://arxiv.org/abs/hep-ph/0005122}{{\ttfamily hep-ph/0005122}}].

\bibitem{Aoki:2000wm}
K.~Aoki, \emph{{Introduction to the nonperturbative renormalization group and
  its recent applications}},
  \href{https://doi.org/10.1142/S0217979200000923}{\emph{Int.J.Mod.Phys.}
  {\bfseries B14} (2000) 1249}.

\bibitem{Bagnuls:2000ae}
C.~Bagnuls and C.~Bervillier, \emph{{Exact renormalization group equations. An
  Introductory review}},
  \href{https://doi.org/10.1016/S0370-1573(00)00137-X}{\emph{Phys. Rept.}
  {\bfseries 348} (2001) 91}
  [\href{https://arxiv.org/abs/hep-th/0002034}{{\ttfamily hep-th/0002034}}].

\bibitem{Polonyi:2001se}
J.~Polonyi, \emph{{Lectures on the functional renormalization group method}},
  \href{https://doi.org/10.2478/BF02475552}{\emph{Central Eur. J. Phys.}
  {\bfseries 1} (2003) 1}
  [\href{https://arxiv.org/abs/hep-th/0110026}{{\ttfamily hep-th/0110026}}].

\bibitem{Pawlowski:2005xe}
J.~M. Pawlowski, \emph{{Aspects of the functional renormalisation group}},
  \href{https://doi.org/10.1016/j.aop.2007.01.007}{\emph{Annals Phys.}
  {\bfseries 322} (2007) 2831}
  [\href{https://arxiv.org/abs/hep-th/0512261}{{\ttfamily hep-th/0512261}}].

\bibitem{Gies:2006wv}
H.~Gies, \emph{{Introduction to the functional RG and applications to gauge
  theories}},
  \href{https://doi.org/10.1007/978-3-642-27320-9_6}{\emph{Lect.Notes Phys.}
  {\bfseries 852} (2012) 287}
  [\href{https://arxiv.org/abs/hep-ph/0611146}{{\ttfamily hep-ph/0611146}}].

\bibitem{Delamotte:2007pf}
B.~Delamotte, \emph{{An Introduction to the nonperturbative renormalization
  group}}, \href{https://doi.org/10.1007/978-3-642-27320-9_2}{\emph{Lect. Notes
  Phys.} {\bfseries 852} (2012) 49}
  [\href{https://arxiv.org/abs/cond-mat/0702365}{{\ttfamily
  cond-mat/0702365}}].

\bibitem{Sonoda:2007av}
H.~Sonoda, \emph{{The Exact Renormalization Group: Renormalization theory
  revisited}},  9, 2007, \href{https://arxiv.org/abs/0710.1662}{{\ttfamily
  0710.1662}}.

\bibitem{Igarashi:2009tj}
Y.~Igarashi, K.~Itoh and H.~Sonoda, \emph{{Realization of Symmetry in the ERG
  Approach to Quantum Field Theory}},
  \href{https://doi.org/10.1143/PTPS.181.1}{\emph{Prog. Theor. Phys. Suppl.}
  {\bfseries 181} (2010) 1} [\href{https://arxiv.org/abs/0909.0327}{{\ttfamily
  0909.0327}}].

\bibitem{Rosten:2010vm}
O.~J. Rosten, \emph{{Fundamentals of the Exact Renormalization Group}},
  \href{https://doi.org/10.1016/j.physrep.2011.12.003}{\emph{Phys. Rept.}
  {\bfseries 511} (2012) 177}
  [\href{https://arxiv.org/abs/1003.1366}{{\ttfamily 1003.1366}}].

\bibitem{Litim:2001up}
D.~F. Litim, \emph{{Optimized renormalization group flows}},
  \href{https://doi.org/10.1103/PhysRevD.64.105007}{\emph{Phys. Rev.}
  {\bfseries D64} (2001) 105007}
  [\href{https://arxiv.org/abs/hep-th/0103195}{{\ttfamily hep-th/0103195}}].

\bibitem{Salmhofer:2001tr}
M.~Salmhofer and C.~Honerkamp, \emph{{Fermionic renormalization group flows:
  Technique and theory}}, \href{https://doi.org/10.1143/PTP.105.1}{\emph{Prog.
  Theor. Phys.} {\bfseries 105} (2001) 1}.

\bibitem{Kopietz:2010zz}
P.~Kopietz, L.~Bartosch and F.~Sch\"utz, \emph{{Introduction to the functional
  renormalization group}}, vol.~798. Springer Berlin, Heidelberg, 2010,
  \href{https://doi.org/10.1007/978-3-642-05094-7}{10.1007/978-3-642-05094-7}.

\bibitem{Braun:2011pp}
J.~Braun, \emph{{Fermion Interactions and Universal Behavior in Strongly
  Interacting Theories}},
  \href{https://doi.org/10.1088/0954-3899/39/3/033001}{\emph{J. Phys.}
  {\bfseries G39} (2012) 033001}
  [\href{https://arxiv.org/abs/1108.4449}{{\ttfamily 1108.4449}}].

\bibitem{Metzner:2011cw}
W.~Metzner, M.~Salmhofer, C.~Honerkamp, V.~Meden and K.~Schonhammer,
  \emph{{Functional renormalization group approach to correlated fermion
  systems}}, \href{https://doi.org/10.1103/RevModPhys.84.299}{\emph{Rev. Mod.
  Phys.} {\bfseries 84} (2012) 299}
  [\href{https://arxiv.org/abs/1105.5289}{{\ttfamily 1105.5289}}].

\bibitem{Aoki:2009zza}
K.-I. Aoki and K.~Miyashita, \emph{{Evaluation of the spontaneous chiral
  symmetry breaking scale in general gauge theories with non-perturbative
  renormalization group}},
  \href{https://doi.org/10.1143/PTP.121.875}{\emph{Prog. Theor. Phys.}
  {\bfseries 121} (2009) 875}.

\bibitem{Braun:2017srn}
J.~Braun, M.~Leonhardt and M.~Pospiech, \emph{{Fierz-complete NJL model study:
  Fixed points and phase structure at finite temperature and density}},
  \href{https://doi.org/10.1103/PhysRevD.96.076003}{\emph{Phys. Rev. D}
  {\bfseries 96} (2017) 076003}
  [\href{https://arxiv.org/abs/1705.00074}{{\ttfamily 1705.00074}}].

\bibitem{Braun:2018bik}
J.~Braun, M.~Leonhardt and M.~Pospiech, \emph{{Fierz-complete NJL model study.
  II. Toward the fixed-point and phase structure of hot and dense two-flavor
  QCD}}, \href{https://doi.org/10.1103/PhysRevD.97.076010}{\emph{Phys. Rev. D}
  {\bfseries 97} (2018) 076010}
  [\href{https://arxiv.org/abs/1801.08338}{{\ttfamily 1801.08338}}].

\bibitem{Braun:2019aow}
J.~Braun, M.~Leonhardt and M.~Pospiech, \emph{{Fierz-complete NJL model study
  III: Emergence from quark-gluon dynamics}},
  \href{https://doi.org/10.1103/PhysRevD.101.036004}{\emph{Phys. Rev. D}
  {\bfseries 101} (2020) 036004}
  [\href{https://arxiv.org/abs/1909.06298}{{\ttfamily 1909.06298}}].

\bibitem{Aoki:1999dv}
K.-I. Aoki, K.~Morikawa, J.-I. Sumi, H.~Terao and M.~Tomoyose, \emph{{Wilson
  renormalization group equations for the critical dynamics of chiral
  symmetry}}, \href{https://doi.org/10.1143/PTP.102.1151}{\emph{Prog. Theor.
  Phys.} {\bfseries 102} (1999) 1151}
  [\href{https://arxiv.org/abs/hep-th/9908042}{{\ttfamily hep-th/9908042}}].

\bibitem{Aoki:2014ola}
K.-I. Aoki, S.-I. Kumamoto and D.~Sato, \emph{{Weak solution of the
  non-perturbative renormalization group equation to describe dynamical chiral
  symmetry breaking}}, \href{https://doi.org/10.1093/ptep/ptu039}{\emph{PTEP}
  {\bfseries 2014} (2014) 043B05}
  [\href{https://arxiv.org/abs/1403.0174}{{\ttfamily 1403.0174}}].

\bibitem{tHooft:1979rat}
G.~'t~Hooft, \emph{{Naturalness, chiral symmetry, and spontaneous chiral
  symmetry breaking}},
  \href{https://doi.org/10.1007/978-1-4684-7571-5_9}{\emph{NATO Sci. Ser. B}
  {\bfseries 59} (1980) 135}.

\bibitem{Wells:2013tta}
J.~D. Wells, \emph{{The Utility of Naturalness, and how its Application to
  Quantum Electrodynamics envisages the Standard Model and Higgs Boson}},
  \href{https://doi.org/10.1016/j.shpsb.2015.01.002}{\emph{Stud. Hist. Phil.
  Sci. B} {\bfseries 49} (2015) 102}
  [\href{https://arxiv.org/abs/1305.3434}{{\ttfamily 1305.3434}}].

\bibitem{Frank:2003ve}
M.~Frank, M.~Buballa and M.~Oertel, \emph{{Flavor mixing effects on the QCD
  phase diagram at nonvanishing isospin chemical potential: One or two phase
  transitions?}},
  \href{https://doi.org/10.1016/S0370-2693(03)00607-5}{\emph{Phys. Lett. B}
  {\bfseries 562} (2003) 221}
  [\href{https://arxiv.org/abs/hep-ph/0303109}{{\ttfamily hep-ph/0303109}}].

\bibitem{Boer:2008ct}
D.~Boer and J.~K. Boomsma, \emph{{Spontaneous CP-violation in the strong
  interaction at theta = pi}},
  \href{https://doi.org/10.1103/PhysRevD.78.054027}{\emph{Phys. Rev. D}
  {\bfseries 78} (2008) 054027}
  [\href{https://arxiv.org/abs/0806.1669}{{\ttfamily 0806.1669}}].

\bibitem{Boomsma:2009eh}
J.~K. Boomsma and D.~Boer, \emph{{The High temperature CP-restoring phase
  transition at theta = pi}},
  \href{https://doi.org/10.1103/PhysRevD.80.034019}{\emph{Phys. Rev. D}
  {\bfseries 80} (2009) 034019}
  [\href{https://arxiv.org/abs/0905.4660}{{\ttfamily 0905.4660}}].

\bibitem{Sakai:2011gs}
Y.~Sakai, H.~Kouno, T.~Sasaki and M.~Yahiro, \emph{{Theta vacuum effects on QCD
  phase diagram}},
  \href{https://doi.org/10.1016/j.physletb.2011.10.032}{\emph{Phys. Lett. B}
  {\bfseries 705} (2011) 349}
  [\href{https://arxiv.org/abs/1105.0413}{{\ttfamily 1105.0413}}].

\bibitem{Aoki:1999dw}
K.-I. Aoki, K.~Morikawa, J.-I. Sumi, H.~Terao and M.~Tomoyose, \emph{{Analysis
  of the Wilsonian effective potentials in dynamical chiral symmetry
  breaking}}, \href{https://doi.org/10.1103/PhysRevD.61.045008}{\emph{Phys.
  Rev. D} {\bfseries 61} (2000) 045008}
  [\href{https://arxiv.org/abs/hep-th/9908043}{{\ttfamily hep-th/9908043}}].

\bibitem{Gies:2001nw}
H.~Gies and C.~Wetterich, \emph{{Renormalization flow of bound states}},
  \href{https://doi.org/10.1103/PhysRevD.65.065001}{\emph{Phys. Rev. D}
  {\bfseries 65} (2002) 065001}
  [\href{https://arxiv.org/abs/hep-th/0107221}{{\ttfamily hep-th/0107221}}].

\bibitem{Gies:2002hq}
H.~Gies and C.~Wetterich, \emph{{Universality of spontaneous chiral symmetry
  breaking in gauge theories}},
  \href{https://doi.org/10.1103/PhysRevD.69.025001}{\emph{Phys. Rev. D}
  {\bfseries 69} (2004) 025001}
  [\href{https://arxiv.org/abs/hep-th/0209183}{{\ttfamily hep-th/0209183}}].

\bibitem{Floerchinger:2009uf}
S.~Floerchinger and C.~Wetterich, \emph{{Exact flow equation for composite
  operators}},
  \href{https://doi.org/10.1016/j.physletb.2009.09.014}{\emph{Phys. Lett. B}
  {\bfseries 680} (2009) 371}
  [\href{https://arxiv.org/abs/0905.0915}{{\ttfamily 0905.0915}}].

\bibitem{Mitter:2014wpa}
M.~Mitter, J.~M. Pawlowski and N.~Strodthoff, \emph{{Chiral symmetry breaking
  in continuum QCD}},
  \href{https://doi.org/10.1103/PhysRevD.91.054035}{\emph{Phys. Rev. D}
  {\bfseries 91} (2015) 054035}
  [\href{https://arxiv.org/abs/1411.7978}{{\ttfamily 1411.7978}}].

\bibitem{Braun:2014ata}
J.~Braun, L.~Fister, J.~M. Pawlowski and F.~Rennecke, \emph{{From Quarks and
  Gluons to Hadrons: Chiral Symmetry Breaking in Dynamical QCD}},
  \href{https://doi.org/10.1103/PhysRevD.94.034016}{\emph{Phys. Rev. D}
  {\bfseries 94} (2016) 034016}
  [\href{https://arxiv.org/abs/1412.1045}{{\ttfamily 1412.1045}}].

\bibitem{Denz:2019ogb}
T.~Denz, M.~Mitter, J.~M. Pawlowski, C.~Wetterich and M.~Yamada, \emph{{Partial
  bosonization for the two-dimensional Hubbard model}},
  \href{https://doi.org/10.1103/PhysRevB.101.155115}{\emph{Phys. Rev. B}
  {\bfseries 101} (2020) 155115}
  [\href{https://arxiv.org/abs/1910.08300}{{\ttfamily 1910.08300}}].

\bibitem{Aoki:2017rjl}
K.-I. Aoki, S.-I. Kumamoto and M.~Yamada, \emph{{Phase structure of NJL model
  with weak renormalization group}},
  \href{https://doi.org/10.1016/j.nuclphysb.2018.04.005}{\emph{Nucl. Phys. B}
  {\bfseries 931} (2018) 105}
  [\href{https://arxiv.org/abs/1705.03273}{{\ttfamily 1705.03273}}].

\bibitem{Kodama:1999if}
H.~Kodama and J.-I. Sumi, \emph{{Application of nonperturbative renormalization
  group to Nambu-Jona-Lasinio / Gross-Neveu model at finite temperature and
  chemical potential}}, \href{https://doi.org/10.1143/PTP.103.393}{\emph{Prog.
  Theor. Phys.} {\bfseries 103} (2000) 393}
  [\href{https://arxiv.org/abs/hep-th/9912215}{{\ttfamily hep-th/9912215}}].

\bibitem{Aoki:2015hsa}
K.-I. Aoki and M.~Yamada, \emph{{The RG flow of Nambu\textendash{}Jona-Lasinio
  model at finite temperature and density}},
  \href{https://doi.org/10.1142/S0217751X15501808}{\emph{Int. J. Mod. Phys. A}
  {\bfseries 30} (2015) 1550180}
  [\href{https://arxiv.org/abs/1504.00749}{{\ttfamily 1504.00749}}].

\bibitem{Alexandrou:2020mds}
C.~Alexandrou, A.~Athenodorou, K.~Hadjiyiannakou and A.~Todaro, \emph{{Neutron
  electric dipole moment using lattice QCD simulations at the physical point}},
  \href{https://doi.org/10.1103/PhysRevD.103.054501}{\emph{Phys. Rev. D}
  {\bfseries 103} (2021) 054501}
  [\href{https://arxiv.org/abs/2011.01084}{{\ttfamily 2011.01084}}].

\end{thebibliography}\endgroup
\end{document}